%% file: downscaling.tex
\documentclass[sn-nature]{sn-jnl}

\usepackage{graphicx}%
\usepackage{multirow}%
\usepackage{amsmath,amssymb,amsfonts}%
\usepackage{amsthm}%
\usepackage{mathrsfs}%
\usepackage[title]{appendix}%
\usepackage{xcolor}%
\usepackage{textcomp}%
\usepackage{manyfoot}%
\usepackage{booktabs}%
\usepackage{algorithm}%
\usepackage{algorithmicx}%
\usepackage{algpseudocode}%
\usepackage{listings}%

\graphicspath{{figs/}}
\input{common/math_definition}

\input{common/personal_commands}
\usepackage{soul}

\usepackage{geometry}
\usepackage[T1]{fontenc}
\usepackage{mathptmx}
\usepackage{caption}

\usepackage[parfill]{parskip}
\usepackage{hyperref}
\usepackage{lineno}
\begin{document}

\title[DGD]{Dynamical-generative downscaling of climate model ensembles}


\author*[1]{\fnm{Ignacio} \sur{Lopez-Gomez}}\email{ilopezgp@google.com}

\author[1]{\fnm{Zhong Yi} \sur{Wan}}

\author[1]{\fnm{Leonardo}\sur{Zepeda-Núñez}}

\author[1, 2]{\fnm{Tapio} \sur{Schneider}}

\author[1]{\fnm{John} \sur{Anderson}}

\author*[1]{\fnm{Fei} \sur{Sha}}\email{fsha@google.com}

\affil[1]{\orgdiv{Google Research}, \orgaddress{\city{Mountain View}, \state{CA}, \country{USA}}}
\affil[2]{\orgdiv{California Institute of Technology}, \orgaddress{\city{Pasadena}, \state{CA}, \country{USA}}}



\maketitle
\input{nc_text/main_abstract}
\input{nc_text/main_significance_statement}
\newpage
\input{nc_text/main_intro}

\input{nc_text/main_results}
\input{nc_text/main_discuss}

\input{nc_text/main_method}

\section*{Acknowledgement}

We thank Alex Hall and Stefan Rahimi for preliminary discussions and for providing information about the WUS-D3 dataset. We also thank Rob Carver for initial discussions about the data, and Lizao Li and Stephan Hoyer for insightful feedback.


\bibliography{common/downscaling}

\newpage
\resetlinenumber[719]
\renewcommand{\appendixpagename}{Supplementary Material}
\appendix
\appendixpage
\renewcommand{\thesection}{S\arabic{section}}
\setcounter{section}{0}
\renewcommand{\thefigure}{S\arabic{figure}}    
\setcounter{figure}{0}
\renewcommand{\theequation}{S\arabic{equation}}
\setcounter{equation}{0}
\captionsetup[figure]{labelfont=bf, labelsep=period}
\setcounter{page}{1}

\input{nc_text/suppl_method_detail}
\input{nc_text/suppl_results_detail}
\end{document}

%% file: common/math_definition.tex
\usepackage{amssymb}
\usepackage{amsmath,amsfonts}
\usepackage{amsopn,bm,mathtools}

\usepackage{nicefrac}       

\newcommand{\ProbOpr}[1]{\mathbb{#1}}

\newcommand{\expect}[2]{%
\ifthenelse{\equal{#2}{}}{\ProbOpr{E}_{#1}}
{\ifthenelse{\equal{#1}{}}{\ProbOpr{E}\left[#2\right]}{\ProbOpr{E}_{#1}\left[#2\right]}}} 
\newcommand{\var}[2]{%
\ifthenelse{\equal{#2}{}}{\ProbOpr{VAR}_{#1}}
{\ifthenelse{\equal{#1}{}}{\ProbOpr{VAR}\left[#2\right]}{\ProbOpr{VAR}_{#1}\left[#2\right]}}} 



%% file: common/personal_commands.tex
\usepackage{xspace}

\let\olditemize=\itemize
\let\endolditemize=\enditemize
\renewenvironment{itemize}{
    \olditemize
    \setlength{\itemsep}{0pt}
    \setlength{\parskip}{0pt}
}{\endolditemize}

\let\oldenumerate=\enumerate
\let\endoldenumerate=\endenumerate
\renewenvironment{enumerate}{
    \oldenumerate
    \setlength{\itemsep}{0pt}
    \setlength{\parskip}{0pt}
}{\endoldenumerate}

\makeatletter
\DeclareRobustCommand\onedot{\futurelet\@let@token\@onedot}
\def\@onedot{\ifx\@let@token.\else.\null\fi\xspace}

\makeatother

\usepackage[colorinlistoftodos,linecolor=blue,backgroundcolor=white,bordercolor=blue,textsize=tiny,textwidth=28mm]
{todonotes}
\newcounter{task}
\newcounter{fs}
\newcounter{larry}
\newcounter{ig}
\newcounter{rc}
\newcommand{\eat}[1]{}

\usepackage{xcolor}
\usepackage{soul}
\sethlcolor{yellow}

%% file: nc_text/main_abstract.tex
\begin{abstract}

Regional high-resolution climate projections are crucial for many applications, such as agriculture, hydrology, and natural hazard risk assessment. Dynamical downscaling, the state-of-the-art method to produce localized future climate information, involves running a regional climate model (RCM) driven by an Earth System Model (ESM), but it is too computationally expensive to apply to large climate projection ensembles. We propose a novel approach combining dynamical downscaling with generative artificial intelligence to reduce the cost and improve the uncertainty estimates of downscaled climate projections. In our framework, an RCM dynamically downscales ESM output to an intermediate resolution, followed by a generative diffusion model that further refines the resolution to the target scale.
This approach leverages the generalizability of physics-based models and the sampling efficiency of diffusion models, enabling the downscaling of large multi-model ensembles.
We evaluate our method against dynamically-downscaled climate projections from the CMIP6 ensemble. Our results demonstrate its ability to provide more accurate uncertainty bounds on future regional climate than alternatives such as dynamical downscaling of smaller ensembles, or traditional empirical statistical downscaling methods. We also show that dynamical-generative downscaling results in significantly lower errors than bias correction and spatial disaggregation (BCSD), and captures more accurately the spectra and multivariate correlations of meteorological fields. These characteristics make the dynamical-generative framework a flexible, accurate, and efficient way to downscale large ensembles of climate projections, currently out of reach for pure dynamical downscaling.
\end{abstract}

%% file: nc_text/main_significance_statement.tex
\section*{Significance statement}

Assessments of climate risk at a regional level are a crucial source of information for climate resilience and adaptation policies. The current regional climate modeling paradigm, which leverages physics-based models to downscale climate projections over limited areas, is too costly to apply to the large climate projection ensembles that are now becoming available. This hinders our ability to capture accurately the uncertainty in regional climate projections. We propose an alternative paradigm that jointly exploits physics-based models and generative artificial intelligence to drastically reduce the cost of downscaling climate projections, while retaining the skill of physics-based approaches. This framework enables translating large climate projection ensembles into impact-relevant climate risk assessments.

%% file: nc_text/main_intro.tex
\section{Introduction}
\label{sIntro}

Regional climate projections below the 10 km scale represent a valuable source of information to stakeholders in need of climate risk assessments. Higher-resolution projections enable a more faithful representation of orography \citep{Pepin2015}, land-atmosphere interactions \citep{Seneviratne2006}, and mesoscale convective systems \citep{Kendon2014}, all of which greatly influence the magnitude and frequency of local extreme weather events \cite{Diffenbaugh2005, Gutowski2020}. Localized climate data is particularly necessary in coastal and mountainous regions, where landscape changes at the kilometer scale largely shape the local climatology \cite{Zangl2005, Hughes2010, Steele2015}. Sectors in need of this kind of granular information include agriculture \citep{Wang2020}, hydrology \cite{Teutschbein2012}, energy \citep{Khan2021}, and natural hazard risk assessment \citep{Knutson2013, Goss2020}.

The need for high-resolution data has prompted the creation of downscaling frameworks, where a statistical or dynamical model refines the projections provided by a coarser-resolution ESM over an area of interest \citep{Giorgi2019}.
Projecting regional climate change through downscaling is not without caveats: biases in the driving ESM can be exacerbated by the regional model –- the so-called “garbage in, garbage out” problem \cite{Hall2014, Giorgi2019}. Nevertheless, techniques to alleviate such biases are maturing \cite{Bruyere2014, Risser2024}, and downscaling remains the best source of climate data at impact-relevant kilometer scales \citep{pcast_2023}.
Dynamical downscaling, in which a high-resolution regional climate model (RCM) is driven with the large-scale and boundary conditions from a global model, is widely recognized as the state-of-the-art method to obtain regional information about future climates \citep{ipcc_ar5_regional}.
Dynamical downscaling is often performed in nested stages, each one increasing the resolution and reducing the spatial extent of the simulated domain \citep{Kurihara1998, Gao2006, Mahoney2012, Huang2020}. While dynamical downscaling enables trading spatial extent for resolution, it remains a computationally intensive task: a computational budget sufficient to simulate the global climate at 100 km resolution would fall short of that required to downscale a region the size of Spain to 10 km resolution. Computational constraints also limit the number of global projections that can be dynamically downscaled, resulting in data coverage gaps and an incomplete assessment of model uncertainty and internal variability in regional climate projections \cite{Pierce2009, Goldenson2023}.

\begin{figure}[h]
    \centering
    \includegraphics[width=\columnwidth,draft=false]{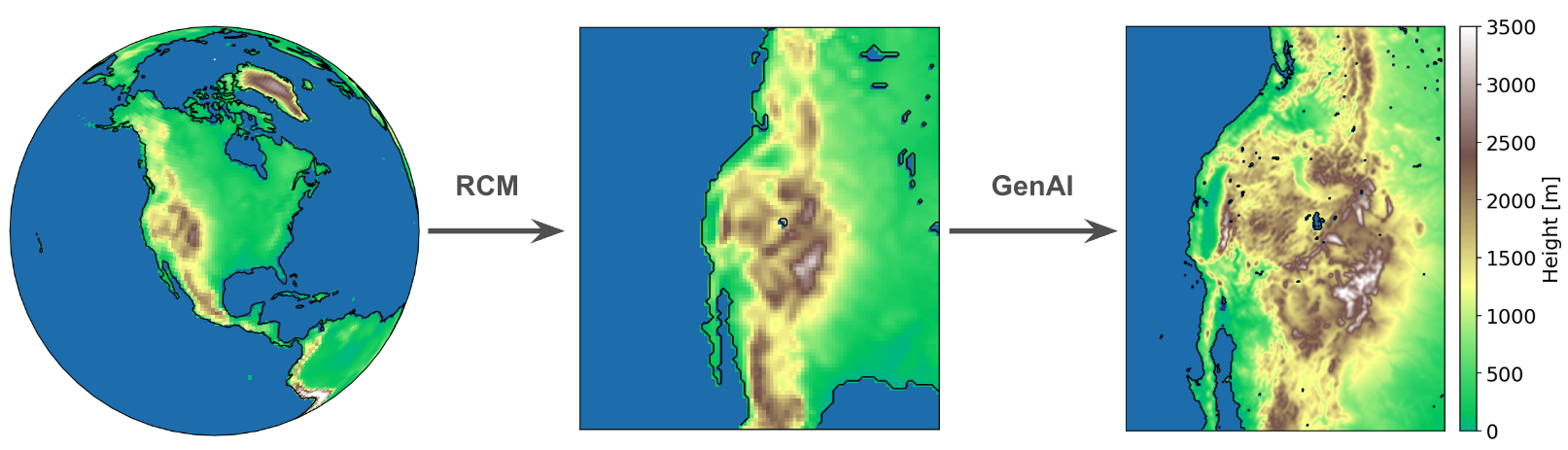}  
    \caption{{Schematic of the dynamical-generative downscaling framework.} A regional climate model (RCM) is used to downscale global simulations from different ESMs to an intermediate grid. A generative artificial intelligence system (GenAI), such as a diffusion model, is then used to further downscale the RCM output to the desired resolution. The topographic height is shown at 100 km (left), 45 km (middle), and 9 km (right) resolution, to showcase landscape changes at the different scales of the process. Water bodies are highlighted in blue.
    \label{fig:topography}}
\end{figure}

To address these issues, we propose to leverage generative artificial intelligence jointly with dynamical downscaling to make downscaling multi-model climate projection ensembles feasible.
In our proposed framework, sketched in Figure \ref{fig:topography}, an RCM is first used to downscale ESM output to an intermediate, but still coarse and inexpensive, resolution. In a second stage, a probabilistic diffusion model is used to efficiently downscale the intermediate RCM fields to the target resolution. Notably, the first dynamical downscaling stage maps input ESM data to atmospheric states that are consistent with the dynamics, resolution, and subgrid-scale parameterizations of a single RCM. This alleviates the need of the generative stage to generalize across the wide range of resolutions and physics of independent ESMs, and enables us to learn a generative model capable of downscaling multi-model ensembles using as training data dynamically downscaled output from a single ESM.

The output from the first stage is then fed into a generative diffusion model, which completes the downscaling process inexpensively, making efficient use of modern accelerators.
In addition, the generative stage
enables sampling the uncertainty of the high-resolution fields given their large-scale context, retaining variable and spatial correlations that are crucial to assess the likelihood of compound extreme events \citep{Kornhuber2023, Anderson2024}.
The ability to capture the multivariate uncertainty of the downscaling process makes generative downscaling an ideal strategy for the second stage of our framework, compared to other statistical or deterministic methods \citep{Sha2020, quesada_chacon_2023, Soares2024}.
Altogether, dynamical-generative downscaling enables obtaining regional climate projections with more accurate uncertainty bounds than those afforded by statistical downscaling methods or computationally limited dynamical downscaling approaches.

Our methodology is inspired by the remarkable ability of probabilistic diffusion models to perform conditional sampling of high-dimensional meteorological fields \citep{Li2024, mardani2023, price2024, Wan_2023}. Prior work has established the utility of generative models for downscaling short-time precipitation forecasts using radar data \citep{Harris2022-af}, and for downscaling historical weather reanalyses \citep{mardani2023, Ling2024}. Here we demonstrate how probabilistic diffusion models trained on data from a single ESM can be used to downscale multi-model climate projections, relying on the proven generalization abilities of physical RCMs. This task requires generalization to unseen climate forcings and is therefore more challenging than downscaling present weather. We evaluate the merits of our approach using as ground truth dynamically downscaled climate projections from the CMIP6 ensemble.

%% file: nc_text/main_results.tex
\section{Results}
\label{sResults}

We demonstrate our dynamical-generative downscaling framework using hourly output from the recently developed Western United States Dynamically Downscaled Dataset (WUS-D3) \citep{Rahimi2024}. WUS-D3 contains nested downscaled climate projections of a CMIP6 multi-model ensemble over the western United States. The first downscaling stage covers the central domain shown in Figure \ref{fig:topography} along with a $675~\mathrm{km}$ east Pacific extension, at an intermediate resolution of $45~\mathrm{km}$. The second and final stage yields climate projections covering the rightmost domain depicted in Figure \ref{fig:topography} at a resolution of $9~\mathrm{km}$. The geographical diversity of this region cannot be fully captured by coarse-resolution simulations and highlights the need for downscaled climate projections.

The CMIP6 climate projections in WUS-D3 are taken from the ScenarioMIP intercomparison project \citep{ONeill_2016} and follow the anthropogenic forcing conditions specified by the Shared Socioeconomic Pathway 3 (SSP3-7.0) \citep{Riahi2017}. Our goal is to capture the internal variability and model uncertainty of climate projections under this forcing. From the WUS-D3 multi-model ensemble we select 8 ESMs for which historical biases, as well as interpolation errors in the sea surface temperature of the Gulf of California, were removed prior to downscaling (see Methods for more details). The RCM used for dynamical downscaling is the Weather Research and Forecasting Model (WRF), in its version 4.1.3 \citep{skamarock2021}.

We seek to substitute the dynamical downscaling stage from the intermediate 45 km resolution to the final $9~\mathrm{km}$ resolution with a generative model. This dynamical downscaling stage is roughly 40 times more computationally intensive than the first downscaling stage to 45 km resolution, covering about a third of the area with 5 times higher resolution. Therefore, our dynamical-generative downscaling framework is designed to overhaul a component that represents more than 97.5\% of the cost of the original system. To this end, we train a Regional Residual Diffusion-based Downscaling (R2-D2) model to sample the probability density function of the residual between the fine ($9~\mathrm{km}$) and coarse ($45~\mathrm{km}$) resolution values of the meteorological fields specified in Table \ref{tb:fields}. Sampling with R2-D2 is very efficient: using a batch size of 32 samples on 16 NVIDIA A100 GPUs, the generative model can downscale 800 input fields per hour.

\begin{table}[t]
\centering
\caption{List of fields used as inputs and outputs to the generative model. All inputs are taken from the 45 km simulation unless otherwise noted.}
\label{tb:fields}
\begin{tabular}{@{}ll@{}}
\toprule
Field & Type (Resolution) \\
\midrule
Temperature at 2 m & Input, Output \\
Specific humidity at 2 m & Input, Output \\
Zonal wind at 10 m & Input, Output \\
Meridional wind at 10 m & Input, Output \\
Surface pressure & Input, Output \\
Precipitation over last 24 h & Input, Output \\
Precipitation over last 12 h & Input \\
Precipitation over last 6 h & Input \\
Surface downwelling longwave flux & Input \\
Surface upwelling longwave flux & Input \\
Surface downwelling shortwave flux & Input \\
Surface upwelling shortwave flux & Input \\
Surface runoff & Input \\
Snow water equivalent & Input \\
Land mask, terrain height & Input (9 km) \\
Latitude, longitude & Input (9 km) \\
Orographic variance & Input (9 km) \\
\bottomrule
\end{tabular}
\end{table}

The R2-D2 model is a conditional probabilistic diffusion model that samples high-resolution residuals conditioned on the coarse resolution input data, as well as additional static input fields such as the topographic height at the target $9~\mathrm{km}$ resolution. As in previous studies, we target the appropriate residuals in the generative learning task to facilitate learning and greatly improve generalization \citep{mardani2023, Li2024}. The residual modeling approach, combined with the mapping of the parent ESM fields to a common effective resolution provided by WRF in the first dynamical downscaling stage, enables us to train a general generative downscaling model using data from a single CMIP6 model. We present results of dynamical-generative downscaling using an R2-D2 model trained on 80 years of data (2014-2094) from an SSP3-7.0 climate projection simulated by the CanESM5 model \citep{Swart2019}. Model skill is assessed against the full 8-model dynamically downscaled ensemble over the test period 2095-2096. Two baselines are included for comparison: bicubic interpolation of the coarse-resolution fields to the target resolution, which serves as the no-added-value baseline; and the popular Bias Correction and Spatial Disaggregation (BCSD) method \citep{Wood2002, Wood2004}, which serves as a competitive statistical downscaling baseline. Both baselines also take input data from the 45 km simulations, and BCSD shares training data with the R2-D2 model. The results of our analysis are not sensitive to the forcing ESM used to train the generative model, as explored in the Supplementary Material.

\subsection{Generative downscaling skill}
\label{sSkill}

The generative downscaling operator in our framework is inherently probabilistic, since the high-resolution fields in WUS-D3 are only constrained to follow their coarse-resolution counterpart at the domain boundaries. Boundary coupling prevents chaotic divergence of the large-scale fields but allows the finer scales to develop freely \citep{sorensen2024}. We evaluate the probabilistic skill of the R2-D2 model using the continuous ranked probability score (CRPS). The CRPS coincides with the mean absolute error (MAE) in the deterministic limit, which enables us to compare our model with deterministic systems \citep{Hersbach2000}. The Supplementary Material evaluates additional fields and metrics and includes individual generative samples. We generate 32-member ensembles using R2-D2 for each evaluation date in the test dataset.

\begin{figure}[h]
    \centering
    \includegraphics[width=\columnwidth,draft=false]{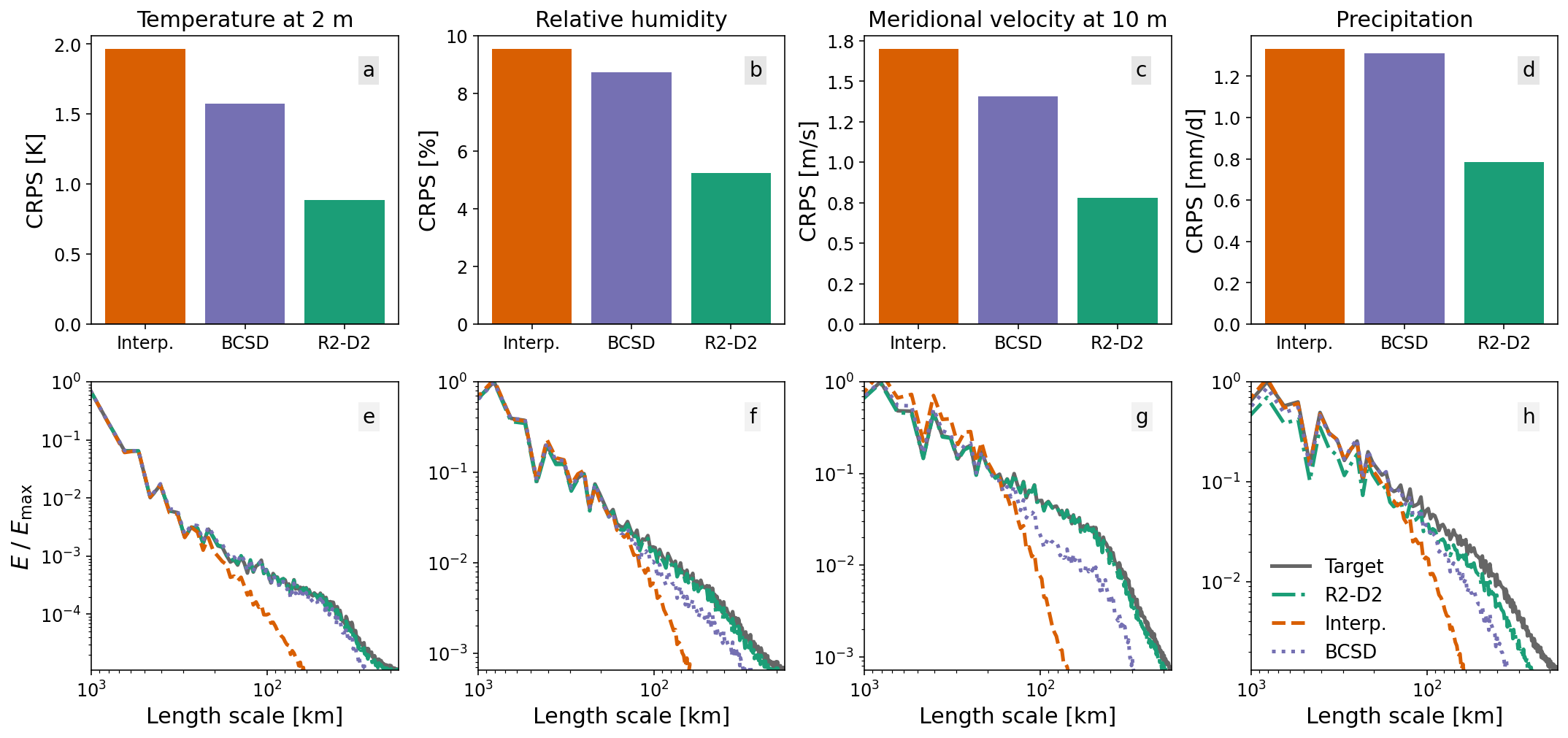}  
    \caption{Land-averaged downscaling skill measured by CRPS (a-d), and radially averaged energy spectra (e-h) for select near-surface meteorological fields.
    Results are computed using 4-hourly data spanning years 2095-2096 of the multi-model SSP3-7.0 climate projection, and shown for cubic interpolation (Interp.), BCSD, and for 32-member R2-D2 ensembles.
    \label{fig:land_avg_CRPS}}
\end{figure}

As shown in Figure \ref{fig:land_avg_CRPS}, the R2-D2 ensembles provide considerable added value over the baselines, reducing the downscaling CRPS with respect to BCSD by more than 40\% for all fields considered.
Error reductions of similar magnitude can be expected for directly modeled fields or derived quantities, since diffusion models retain realistic multivariate correlations. Figure \ref{fig:land_avg_CRPS}b demonstrates this for near-surface relative humidity, diagnosed from the fields directly modeled by R2-D2.  Many downstream application of climate downscaling, such as hydrologic forecasting \citep{burger_2005, maraun_2010}, require their input meteorological fields to have a realistic spatial structure. The radially averaged energy spectra of the generative samples, which characterizes the realism of the downscaled output, follows closely that of the dynamically downscaled fields (Fig. \ref{fig:land_avg_CRPS}e-h). Minor differences are observed for precipitation, but these are still smaller than those observed in the other systems considered.

\begin{figure}[t]
    \centering
    \includegraphics[width=\columnwidth,draft=false]{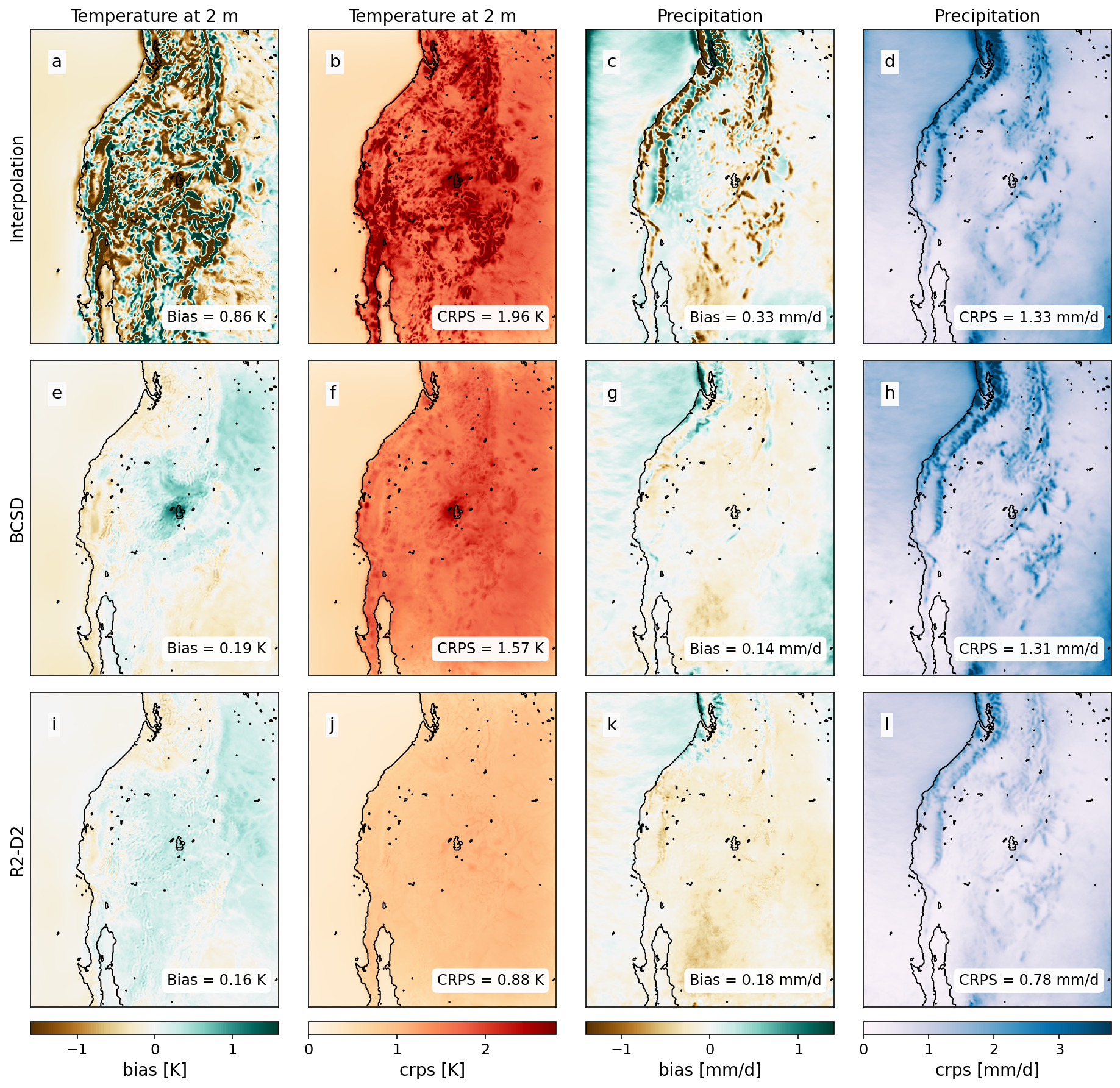}  
    \caption{{Spatial distribution of downscaling bias (first and third columns) and CRPS (second and fourth columns), shown for near-surface temperature and precipitation.}
    {Results are computed using 4-hourly data spanning years 2095-2096 of the multi-model SSP3-7.0 climate projection, and shown for cubic interpolation, BCSD, and for 32-member R2-D2 ensembles. The insets show the land-averaged absolute value of each metric.}
    \label{fig:spatial_bias_and_crps}}
\end{figure}

A fraction of the error reduction afforded by downscaling methods can be attributed to the correction of systematic climatological differences between the coarse and high resolution fields. The extent to which these differences are reduced can be evaluated in terms of the long-term bias, shown in Figure \ref{fig:spatial_bias_and_crps} for near-surface temperature and precipitation. Climatological biases are more profound in regions of complex topography like the Rocky Mountains \citep{Matiu2024} and for meteorological fields with significant lapse rates like temperature, as seen in Figure \ref{fig:spatial_bias_and_crps}a,c for interpolation. Quantile-mapping methods such as BCSD excel at correcting biases by construction, yielding in this case average bias reductions over 58\% for precipitation and 77\% for temperature relative to interpolation.
R2-D2 matches and can even exceed the debiasing skill of quantile-based methods, as demonstrated by the noticeable reduction in near-surface temperature bias near the Great Salt Lake in Figure \ref{fig:spatial_bias_and_crps}i. The ability of R2-D2 to capture multivariate correlations ensures that arbitrary diagnostic variables constructed from the originally modeled fields are also unbiased. The debiasing skill of quantile-based methods such as BCSD degrades for such diagnostics, as shown in the Supplementary Material for near-surface wind speed and relative humidity.

In addition to debiasing, the R2-D2 model further reduces the downscaling error by capturing the distribution of high-resolution anomalies conditioned on the coarse-resolution fields. This amounts to a general reduction of the downscaling error for all fields and over all regions compared to quantile-based methods, as presented in Figure \ref{fig:spatial_bias_and_crps}, and in the Supplementary Material in terms of the root mean square error. The enhanced conditional sampling skill provided by the R2-D2 model is important for applications that focus on particular events, such as extreme event attribution \citep{Emanuel2008,Trenberth2015} and climate storyline analysis \citep{Fischer2023, Liu2024}.

\subsection{Quantifying uncertainty in multi-model climate projections}

Quantifying model uncertainty and internal variability is crucial for accurate regional climate risk assessments but generally unaffordable through pure dynamical downscaling \citep{Pierce2009}. We analyze the ability of dynamical-generative downscaling to capture the full distribution of regional climate projections by downscaling end-of-century multi-model projections under the SSP3-7.0 forcing scenario. Reliable estimates of the regional projection quantiles are particularly important, as they directly impact the accuracy of climate risk forecasts.

\begin{figure}[t]
    \centering
    \includegraphics[width=0.9\columnwidth,draft=false]{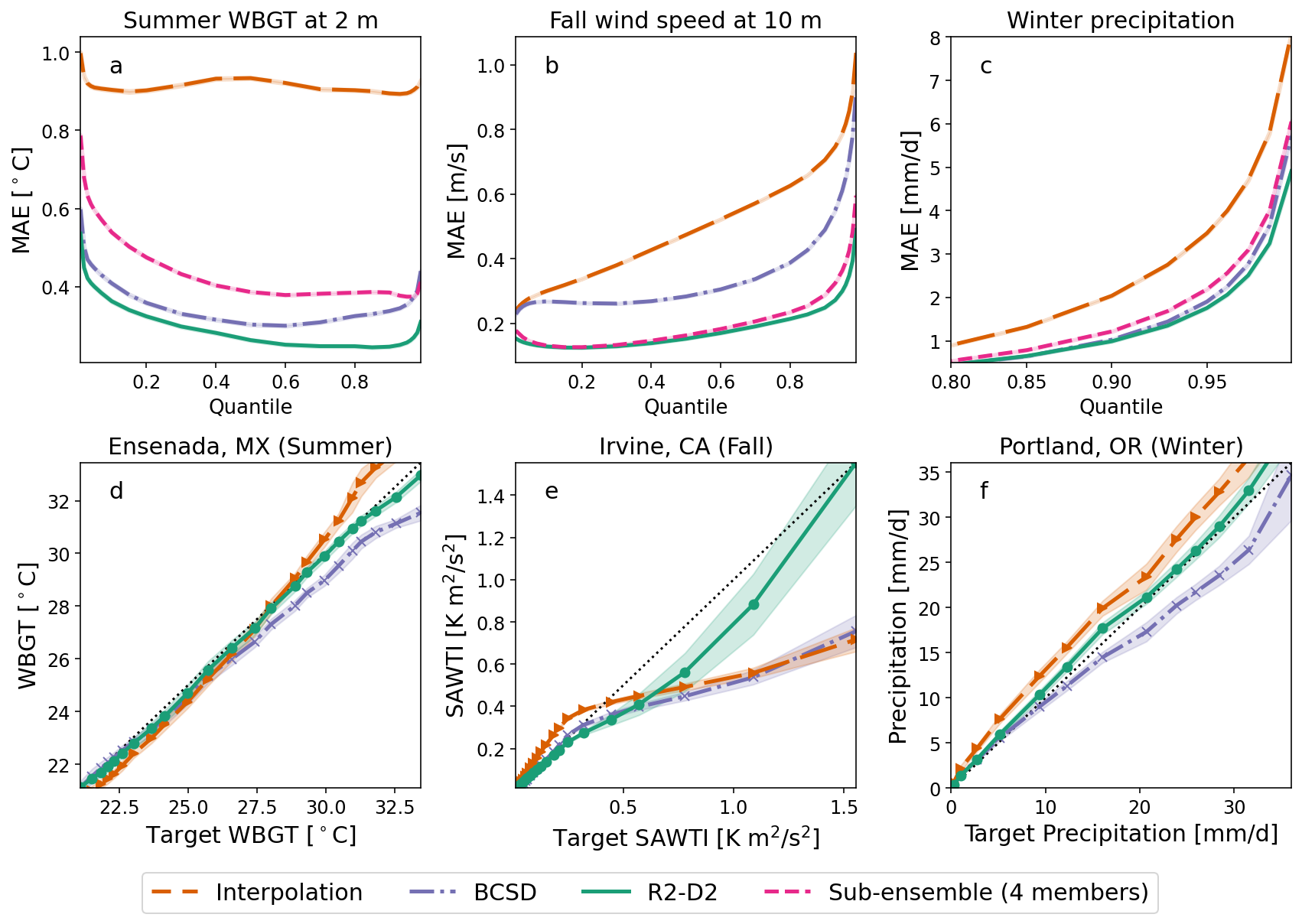}  
    \caption{{Assessment of downscaled multi-model ensemble distribution fidelity.} Top: MAE of downscaled quantiles over land, with respect to the quantiles of the full dynamically downscaled ensemble. Bottom: Quantile-quantile plots at specific locations with respect to the full dynamically downscaled ensemble.
    {Quantiles from 0.01 to 0.99 are computed using daily snapshots at 00 UTC for three-month periods of 2095. Results are shown for cubic interpolation, BCSD, the generative model R2-D2, and for the average over 4-member sub-ensembles. Uncertainty estimates represent the bootstrapped sample standard deviation.}
    \label{fig:quantiles}}
\end{figure}

Figure \ref{fig:quantiles}a-c evaluates the land-averaged mean absolute error (MAE) of the multi-model projection quantiles predicted by the dynamical-generative framework for selected variables and seasons, with respect to the full dynamically downscaled multi-model ensemble. Quantiles of the multi-model ensemble are computed using daily snapshots from all ESMs over the three-month periods of June-August (Summer), September-November (Fall), and December-February (Winter). The daily frequency is chosen so as to prevent the diurnal cycle from dominating the projection distributions in variables such as temperature. Therefore, the distributions shown capture the spread from internal variability and model uncertainty, and the lowest and highest quantiles represent extreme conditions with respect to the seasonal climatology.

To demonstrate the added value of dynamical-generative downscaling, we compare our framework to an alternative approach that dynamically downscales a smaller ensemble of climate projections to the target resolution. This baseline is representative of the current practice of model selection under computational constraints \citep{Goldenson2023}. We show results for 4-member sub-ensembles, or sub-ensembles with half the size of the full ensemble. This alternative is significantly more expensive than our hybrid approach, which only requires dynamical downscaling of one ESM to the target 9 km resolution in order to train the generative model.

Figure \ref{fig:quantiles}a shows results for the summer near-surface wet-bulb globe temperature (WBGT), computed following the simplified definition used by the Australian Bureau of Meteorology \citep{Willett2012}; this and all other derived variables are defined in the Supplementary Material. The WBGT provides a measure of heat stress that accounts for both temperature and humidity. As such, the ability to capture its summer daytime distribution, and its upper quantiles in particular, is important for extreme heat risk assessment. The dynamical-generative approach outperforms all other baselines in this task, including the smaller dynamically-downscaled ensembles. Its downscaled climate projection matches the target distribution better at all quantiles, reducing the $99\%$-quantile error by over $25\%$ with respect to the 4-member sub-ensembles.

Quantile errors are also shown for fall near-surface wind speed and winter precipitation in Figure \ref{fig:quantiles}b-c. The proposed framework yields $99\%$-quantile error reductions of over $17\%$ for wind speed and $14\%$ for precipitation with respect to the best performing baseline, demonstrating the consistently superior performance of R2-D2 across seasons and downscaled fields. Results for additional variables and spring can be found in the Supplementary Material. The difference between dynamical-generative downscaling and sub-ensemble downscaling is more substantial for temperature and precipitation than it is for wind speed. In the first two cases, dynamical-statistical downscaling aided by BCSD can also outperform pure dynamical downscaling of smaller ensembles, because model uncertainty is an important contributor to future climate uncertainty. However, for variables that show a less pronounced intermodel spread contribution to the total uncertainty, such as wind speed, pure dynamical downscaling of sub-ensembles is superior to BCSD. The dynamical-generative approach improves upon the baseline approaches in all cases, particularly for extreme quantiles.

Local future climatological distributions projected with our framework are compared to the target distribution and the baselines in Figure \ref{fig:quantiles}d-f for selected regions. Summer temperatures in the coastal city of Ensenada, in Baja California, are strongly affected by the inland extent of marine stratocumulus cloud decks \citep{Clemesha2016}. This makes Ensenada a location where downscaling can add important value in assessments of extreme heat risk. As shown in Figure \ref{fig:quantiles}d, BCSD is able to correct the distribution for low and moderate quantiles, but underestimates the probability of hot extremes compared to the target or R2-D2. The generative model is able to track well the quantiles projected by the full dynamically-downscaled ensemble, even at the tails of the distribution.

Downscaled projections are similarly essential for fire risk assessment in Southern California, where the hot and dry Santa Ana winds can fan wildfires in the fall, before the onset of the rain season \citep{Hughes2010}. The wildfire potential near Irvine (California) is explored in Figure \ref{fig:quantiles}e in terms of the weather component of the Santa Ana Wildfire Threat Index (SAWTI; \cite{Rolinski2016}), which is proportional to the kinetic energy of the near-surface winds and the dewpoint deficit. In this case, R2-D2 is able to capture accurately the risk of the conditions most conducive to wildfires, characterized by high values of SAWTI. Both interpolation and BCSD strongly underestimate the future wildfire risk. Finally, results are also shown for winter precipitation over Portland, Oregon in Figure \ref{fig:quantiles}f. In this case as well, the climatological distribution projected by the dynamical-generative framework closely follows that of the full dynamically downscaled climate ensemble, at a small fraction of the cost.

\subsection{Regional analysis of compound extremes}

\begin{figure}[h]
    \centering
    \includegraphics[width=0.9\columnwidth,draft=false]{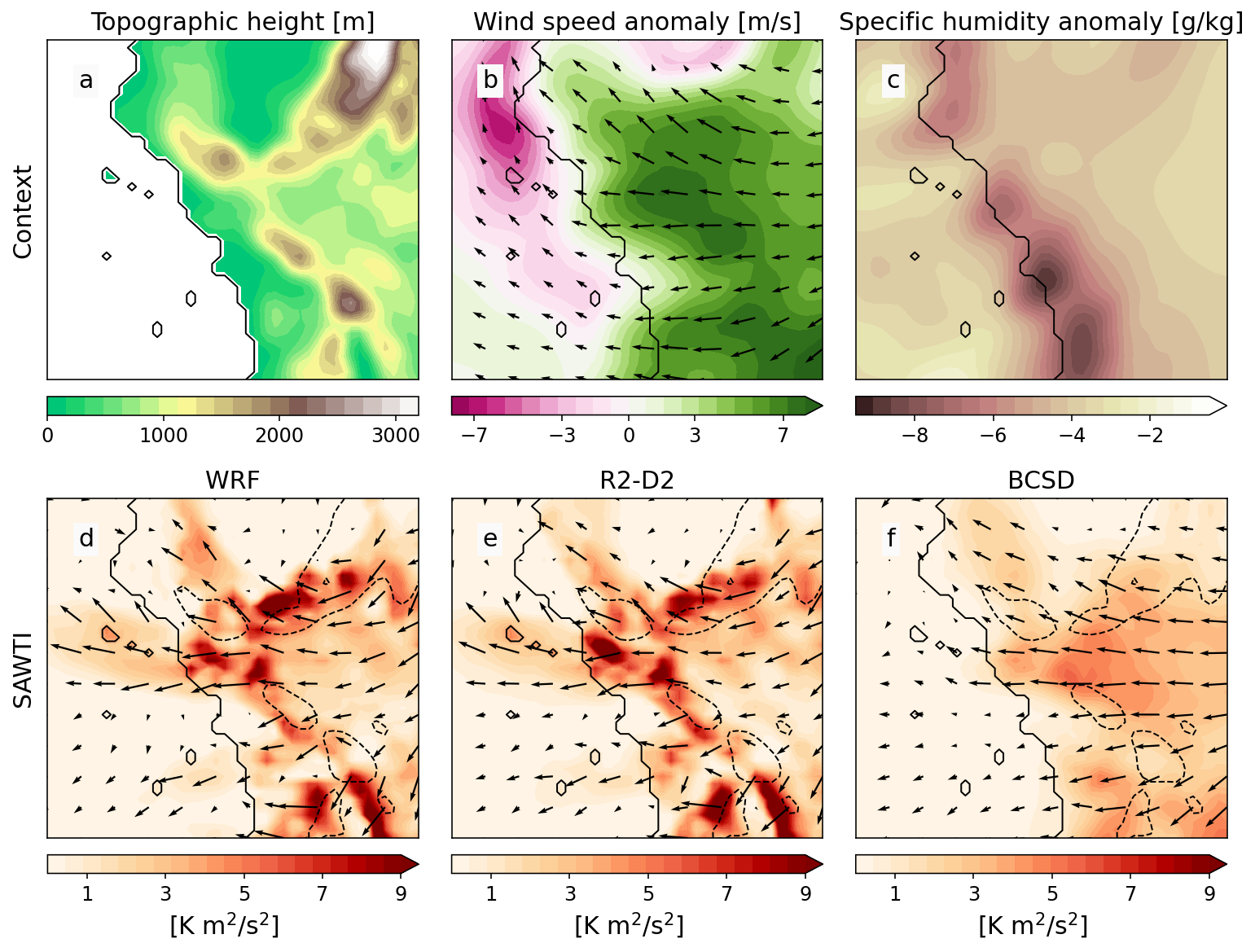}  
    \caption{{Analysis of the strongest Santa Ana wind event in the multi-model projection for the period September-November 2095.} Top: Topographic and coarse-resolution context of the event. The anomaly fields (b,c) show the 45-km resolution anomalies with respect to their September-November 2095 climatology. Bottom: Downscaled Santa Ana wildfire threat index (SAWTI) from WRF (d), R2-D2 (e), and BCSD (f). Quiver plots represent the magnitude and direction of 10 m winds, and dashed contours represent the 1200 m isohypse of the mountain ranges.}
    \label{fig:santa_ana}
\end{figure}

We further demonstrate the ability of dynamical-generative downscaling to capture regional compound extremes by analyzing the strongest Santa Ana wind event projected by the dynamically downscaled multi-model ensemble over the period September-November 2095. The event is selected as the 00 UTC snapshot with the highest average SAWTI over Southern California in the dynamically downscaled ensemble. This Santa Ana wind event occurred on November 13, 2095 of the SSP3-7.0 projection corresponding to the forcing model EC-Earth3-Veg \citep{Doscher2022}. Figure \ref{fig:santa_ana}b,c depict the 45-km resolution conditions during this event in terms of their anomalies with respect to the September-November 2095 climatology. The conditions for this date were characterized by stronger than usual northeasterly winds, and anomalously dry air, particularly close to the coast.

High spatial resolution is crucial to resolve the local acceleration of Santa Ana winds as they make their way from the Mojave desert to the coast through the mountain passes, depicted in Figure \ref{fig:santa_ana}a. The wind flow and SAWTI values of the dynamically downscaled simulation are shown in Figure \ref{fig:santa_ana}d. The 9-km WRF simulations capture the deflection of the flow by the mountain ranges and the intensification of wildfire risk downstream, as the hot and dry desert air descends into the valleys and coastal areas of Southern California. R2-D2 captures these patterns as well, projecting similarly strong SAWTI conditions along the Santa Clara River valley and extending offshore into the Channel Islands. In contrast, BCSD is unable to capture the magnitude and spatial structure of wildfire risk, due to its inability to map coarse-resolution conditions to high-resolution climatological anomalies.

Another benefit of R2-D2 is its ability to assess the extent to which the strong wildfire conditions are determined by the coarse-resolution context. This can be analyzed in terms of the spread of the generative samples conditioned on the same coarse-resolution fields. In this particular case, the SAWTI conditions were tightly controlled by the coarse-resolution input, resulting in low variability in the generated samples and high confidence in severe wildfire risk. Additional samples supporting this tight large-scale control are included in the Supplementary Material, along with an analysis of the spread of the generative samples, which typically accounts for more than 70\% of the downscaling error. More generally, the dynamical-generative framework provides estimates of the internal variability of the downscaled fields given a large-scale context, which can be leveraged to construct counterfactuals and study the drivers of regional extremes \citep{Sillmann_2021, Zscheischler2018}.

%% file: nc_text/main_discuss.tex
\section{Discussion}
\label{sDiscuss}

Dynamical-generative downscaling combines the physical basis of dynamical downscaling with the sampling efficiency of diffusion models to provide regional climate projections that capture the full range of scenarios projected by existing ESMs. Dynamically downscaling coarse climate data from a wide range of ESMs to an intermediate resolution yields fields that are physically consistent with the RCM. This greatly simplifies the learning task of the generative model, by reducing the spectrum of conditioning inputs to those consistent with the dynamics of the RCM. The initial dynamical projection also creates a stable basis for incorporating high-resolution details, which facilitates residual learning.

These characteristics are exploited by the diffusion model in the generative stage, which can be trained on dynamically downscaled data from a single ESM to construct an efficient probabilistic sampler of high-resolution meteorological fields. The sampled fields have realistic spectra and capture the multivariate uncertainty inherent in the downscaling process. Capturing spatial and field correlations maximizes the added value of the high-resolution projections by enabling downstream users to derive their own relevant climate indicators without loss of accuracy. Quantile-based methods such as BCSD do not capture these correlations, which can lead to the underestimation of compound risks \citep{Zscheischler2018,  Chandel2024}. Even in the case of univariate climate distributions, the generative R2-D2 model outperforms BCSD in downscaling skill. In addition, downscaled climate risk assessments with the proposed framework capture the uncertainty underlying climate projections better than dynamical downscaling of smaller ensembles, the current paradigm.

The dynamical-generative framework can be designed to be substantially more economical than pure dynamical downscaling, facilitating its application to very large climate model ensembles. In this study, the generative stage substitutes a component that consumes roughly 97.5\% of the total computational budget in the original system. We document the use of the framework to downscale an ensemble of 8 ESMs. In this case, our method saves 85\% of the dynamical downscaling cost---a percentage that would increase for larger climate model ensembles. Inference with the diffusion model is relatively cheap: using a batch size of 32 samples on 16 NVIDIA A100 GPUs, the throughput is 800 samples per hour. Moreover, the generative stage can sample downscaled fields without an initial spin-up time, which further boosts its efficiency compared to physics-based downscaling.

As previously demonstrated in weather forecasting \citep{Li2024}, diffusion models can be used in conjunction with dynamical models of the atmosphere to reduce the cost of ensemble projections and augment the added value of the entire system. In the case of climate downscaling, this enables downscaling much larger climate model ensembles than those currently afforded by the dynamical downscaling paradigm \citep{Gutowski2020, Goldenson2023}. This is essential to capture the regional impacts of climate change as projected by current state-of-the-art climate models.

%% file: nc_text/main_method.tex
\section{Methods}
\label{sMethod}

\subsection{Data for learning and evaluation}

We derive our input and output data from the WUS-D3 dataset \citep{Rahimi2024}. WUS-D3 employs WRF to dynamically downscale global climate projections to an intermediate resolution of 45 km over a region covering the entire western United States, and extensions into the Midwest, the Pacific, western Canada, and northern Mexico, and shown in Figure \ref{fig:topography}b. The 45 km WRF simulations are driven by 6-hourly lateral boundary conditions and nudged toward the large-scale (>1500 km) conditions of the forcing ESM with a relaxation timescale of 1.08 hours. The 45 km grid is subsequently dynamically downscaled to 9 km resolution over the Western
Electricity Coordinating Council (WECC) US coverage area shown in Figure \ref{fig:topography}c, using 6-hourly lateral boundary conditions from the 45 km grid. In this study, we focus on the time-aligned hourly data available from both the coarse and high-resolution grids.

The data used for evaluation comes from future climate projections under the SSP3-7.0 scenario using 8 different ESMs: CanESM5 \citep{Swart2019}, EC-Earth3-Veg \citep{Doscher2022}, UKESM1-0-LL \citep{Sellar2019}, MIROC6 \citep{Tatebe2019}, ACCESS-CM2 \citep{Bi2020}, MPI-ESM1-2-HR \citep{Gutjahr2019}, NorESM2-MM \citep{Seland2020}, and TaiESM1 \citep{Wang2021}.  The output of each ESM is debiased prior to its use as input to WRF \cite{Risser2024}, using as a reference the historical biases with respect to the ERA5 reanalysis \citep{Hersbach2020-iu}. In addition, sea surface temperatures in the Gulf of California are corrected to reflect its observed distribution, unresolved by most ESMs. This correction, as well as the resolution and specific climate projections used, are detailed in depth in the WUS-D3 description paper \citep{Rahimi2024}.

We train a probabilistic diffusion model to conditionally sample hourly snapshots of the meteorological field differences between the 9 km and 45 km resolution simulations. This difference is computed on the 9 km grid, after cubic interpolation of the 45 km resolution data to the 9 km grid. The generative model is conditioned on time-aligned and spatially interpolated 45 km meteorological fields, as well as static information about the 9 km grid. Therefore, all inputs and outputs cover the spatial grid with $340\times270$ degrees of freedom depicted in Figure \ref{fig:topography}. Only spatial downscaling is considered, not time upsampling. To facilitate training, each input and output field is centered and normalized using its temporal and spatial mean and standard deviation in the training set. Further details of our modeling framework are included in the Supplementary Material.

Given a forcing model used for training the generative model (CanESM5 in the main text), the period 2015-2094 is used for training, 2098-2100 for evaluation, and 2095-2097 for testing. Since the generative model only sees one ESM during training and evaluation, all other models used in the evaluation of the framework are part of the test set.

\subsection{Generative model design and training}

The R2-D2 model is a conditional score-based diffusion model \citep{Ho2020} for generative downscaling, with training objective of denoising score matching \citep{vincent2011}. We adopt the denoising formulation introduced by Karras et al. \citep{karras2022elucidating} with a variance-exploding noise schedule. The denoiser is parameterized by a UNet-type convolutional architecture with about 180 million trainable parameters. The diffusion configuration and neural network architecture are described in detail in the Supplementary Material.

We train and sample from the R2-D2 model using classifier-free guidance \citep{ho2022}, with guidance strength of {$g=0.1$} and masking the conditioning input with probability {$p_u=0.1$} during training. We also employ a dropout probability $p_d=0.1$ for the model weights during training. The diffusion model is trained using the Adam optimizer with a batch size of {128 for $2\cdot10^5$} steps, and exponential moving average decay of 0.9999. The learning rate schedule consists of cosine decay after an initial linear ramp-up, with a peak learning rate of {$2\cdot10^{-4}$} and a terminal value of {$1\cdot10^{-6}$}. Training of the model takes about {5.5 days using 16 NVIDIA A100 GPUs}.

\subsection{Baselines}

Dynamical-generative downscaling is compared against three baselines: cubic interpolation, BCSD, and dynamically downscaled sub-ensembles. Cubic interpolation outputs the requested meteorological fields from the 45 km grid, interpolated to the 9 km grid. In terms of the residual modeling approach used in our generative stage, cubic interpolation represents the null or zero residual output. BCSD follows a time-aligned implementation without temporal disaggregation, since inputs and outputs are aligned in our setting \citep{Thrasher2012}. BCSD is performed by debasing the coarse-resolution fields using as a reference the climatology of the low-pass filtered high-resolution fields over the dates and forcing model in the training dataset. After debiasing, spatial disaggregation is performed by retaining the climatological anomalies and substituting the debiased climatological mean by the high-resolution climatological mean. The retained climatological anomalies are additive for all variables, with the exception of precipitation, for which the multiplicative anomalies are retained \citep{Wood2002}. Finally, the climate projection reliability of ESM sub-ensembles is assessed by consid 4-member sub-ensembles containing CanESM5 and three other ESMs. 4 sub-ensembles are evaluated, such that all ESMs are used at least once in the analysis. The reported metrics are the average metrics over the sub-ensembles.

%% file: nc_text/suppl_method_detail.tex
\section{Method details}
\label{sMethodAppendix}

\subsection{Regional Residual Diffusion-based Downscaling}

We consider the problem of obtaining a sampler of high-resolution meteorological fields $\tilde{x} \in \mathbb{R}^{d_1\times d_2 \times n_x}$ over a grid $\Omega$, conditioned on coarse-resolution climate data $y_c$ and high-resolution static information $y_g  \in \mathbb{R}^{d_1\times d_2}$ about the spatial domain discretized by $\Omega$. The coarse-resolution data is assumed to be available over a domain $\Omega_c$, such that $\Omega \subseteq \Omega_c$. This is consistent with the format of nested downscaled climate data. We construct the conditioning tensor $y\in \mathbb{R}^{d_1\times d_2 \times n_y}$ by concatenating static fields $y_g$ with the coarse-resolution fields $y_c$ interpolated to grid $\Omega$. $n_x$ and $n_y$ represent the number of output and input fields, respectively. We seek to obtain a conditional residual sampler of the form
\begin{equation}\label{eq:residual}
    x \sim p(x \vert y), \;\;\; x = \tilde{x} - \tilde{x}_c,
\end{equation}
where $x$ is the difference between the high-resolution fields $\tilde{x}$ and their interpolated coarse-resolution counterparts, denoted by $\tilde{x}_c \in \mathbb{R}^{d_1\times d_2 \times n_x}$. Sampler \eqref{eq:residual} draws samples from the distribution of high-resolution residuals relative to coarse-resolution climate fields consistent with the conditioning data. In our setting, the target fields $\tilde{x}$ are coupled to the coarse-resolution fields $\tilde{x}_c$ through the boundaries. Therefore, the generated samples are designed to add plausible high-resolution details to a given coarse-resolution sample, such that their values coincide at the boundaries.

In this study, the spatial dimensions of the samples are $d_1=340$ and $d_2=270$, the number of conditioning inputs is $n_y=19$, and the model outputs $n_x=6$ high-resolution residual fields. In order to facilitate learning, each input and output field is centered and normalized using its spatial and temporal mean $\mu \in \mathbb{R}$ and standard deviation $s \in \mathbb{R}$ over the entire training dataset.

\subsection{Diffusion model formulation}

Diffusion models gradually add Gaussian noise to samples from a given data distribution until they become indistinguishable from pure noise. Their strength as generative models lies in the fact that the diffusion process can be reversed to enable drawing samples from noise and transforming them back onto the data distribution. The forward diffusion process can be described by the stochastic differential equation (SDE)
\begin{equation}\label{eq:forward_sde}
    dx_t = \sqrt{\dot{\sigma}_t\sigma_t}\; d\omega_t, \quad x_0 \sim p_{\text{data}},
\end{equation}
where $\sigma_t$ is a prescribed noise schedule and a strictly increasing function of the diffusion time $t$, $\dot{\sigma}_t$ denotes its time derivative, and $\omega_t$ is the standard Wiener process. This forward SDE defines the conditional distribution
\begin{equation}\label{eq:perturb_kernel}
    q(x_t|x_0) = \mathcal{N}(x_t; x_0, \sigma^2_tI),
\end{equation}
which is a Gaussian distribution with mean $x_0$ and diagonal covariance $\sigma^2_tI$. For $t = T \gg 0$, we impose $\sigma_T \gg \sigma_{\text{data}}$ such that $q(x_T|x_0) \approx \mathcal{N}(x_T; 0, \sigma^2_TI) = q_T$. Prior work has shown that one can derive a backward SDE, which, when integrated from $T$ to $0$, induces the same marginal distributions $p(x_t)$ as those from the forward SDE~\eqref{eq:forward_sde} \citep{anderson1982reverse, Song_2020}:
\begin{equation}\label{eq:backward_sde}
    dx_t = - 2\dot{\sigma}_t\sigma_t\nabla_{x_t} \log{p\left(x_t, \sigma_t\right)}\;dt + \sqrt{2\dot{\sigma}_t\sigma_t}\;d\omega_t.
\end{equation}
The essence of the diffusion modeling framework lies in representing the score function $\nabla_{x_t} \log{p\left(x_t, \sigma_t\right)}$ with a neural network. Specifically, the neural network parameterization follows Tweedie's formula~\citep{efron2011tweedie}:
\begin{equation}\label{eq:tweedies}
    \nabla_{x_t} \log{p\left(x_t, \sigma_t\right)} = \frac{\mathbb{E}[x_0|x_t]-x_t}{\sigma_t^2} \approx \frac{D_{\theta}(x_t, \sigma_t) - x_t}{\sigma_t^2},
\end{equation}
where $D_{\theta}$ is a denoising network that predicts the denoised data sample $x_0$ given a noisy sample $x_t$. Training $D_{\theta}$ involves sampling both data samples $x_0$ and diffusion times $t$, and optimizing parameters $\theta$ with respect to the mean denoising loss defined by
\begin{equation}\label{eq:dfn_train_loss}
    L(\theta) = \mathbb{E}_{x_0\sim p_{\text{data}}} \mathbb{E}_{t\sim[0, T]} \;\big[w_t\|D_{\theta}(x_0 + \epsilon\sigma_t, \sigma_t) - x_0||^2\big],
\end{equation}
where $\epsilon$ draws from a standard Gaussian $\mathcal{N}(0, I)$ and $w_t$ represents the loss weight for $t$. We use the weighting scheme proposed in~\citep{karras2022elucidating} as well as the pre-conditioning strategies therein to improve training stability.

At sampling time, we substitute the score function in SDE~\eqref{eq:backward_sde} with the learned denoising network $D_\theta$ using expression \eqref{eq:tweedies}. The backward diffusion process is then integrated from $t=T$ to $t=0$, using initial condition $x_T\sim q_T$ and the first order Euler-Maruyama scheme.

In this study, we aim to construct a sampler for the conditional distribution \eqref{eq:residual}, which corresponds to training a denoiser $D_\theta$ that incorporates the condition $y$ as an additional input. We employ classifier-free guidance (CFG)~\citep{ho2022} to find an optimal balance between maintaining coherence with the conditional input and ensuring comprehensive coverage of the target distribution. CFG is implemented through the use of the modified denoising function $\tilde{D}_\theta$ at sampling time,
\begin{equation}\label{eq:guidance}
    \tilde{D}_\theta = (1+g){D}_\theta(x_t, \sigma_t, y) - g{D}_\theta(x_t, \sigma_t, \varnothing),
\end{equation}
where $g$ is the guidance strength and $\varnothing$ is the null conditioning input (i.e., a zero-filled tensor of the same shape as $y$), such that ${D}_\theta(x_t, \sigma_t, \varnothing)$ represents unconditional denoising. The unconditional and conditional denoisers are trained jointly using the same neural network model, by randomly dropping the conditioning input from training samples with probability $p_u$.

\subsection{Model architecture}
\label{sArchitecture}

The diffusion model denoiser $D_\theta$ is implemented using a U-Net architecture \citep{Ronneberger2015}. The denoiser takes as inputs noised samples $x_t$, the conditioning inputs $y$, and the noise level $\sigma_t$. The output is the denoised sample
\begin{equation}
    x'_0 = D_\theta(x_t, \sigma_t, y).
\end{equation}
The denoiser architecture is formed by an encoder, downsampling and upsampling stacks, and a decoder.

\subsubsection{Encoding}

The input noised samples $x_t$ and the conditioning inputs $y$, formed by 2D fields with $340\times270$ spatial degrees of freedom, are first resized to $352\times288$ spatial fields through cubic interpolation to enable multilevel downsampling. Resizing of the noised samples is followed by a dealiasing convolutional layer,
\begin{equation}
    h_x = \text{Conv}_7\circ \mathcal{I}(x_t),
\end{equation}
where $\mathcal{I}$ represents the resizing operator, and $\text{Conv}_k$ denotes a convolutional layer with a $k\times k$ kernel that preserves the number of channels. The conditioning inputs $y$ are passed through a dealiasing convolutional layer with a $3\times3$ kernel, and a Swish nonlinearity,
\begin{equation}
    h_y = \text{Conv}_3^{192}\circ\text{swish}\circ\text{LN}\circ\text{Conv}_3\circ \mathcal{I}(y),
\end{equation}
where $\text{Conv}_k^{N}$ denotes a convolutional layer with $N$ filters, and $\text{LN}$ denotes layer normalization. The resized encodings $h_x$ and $h_y$ are then concatenated in the channel dimension as $h$.

\subsubsection{Downsampling stack}

The downsampling stack takes as input the resized encoding and projects it onto a latent space with $128$ channels through a $3\times3$ convolutional layer,
\begin{equation}\label{eq:init_conv_dstack}
    h_0 = \text{Conv}^{128}_3(h),
\end{equation}
where $h$ is the latent input from the encoding step. The output is passed through 3 downsampling convolutional blocks. Each block halves the degrees of freedom of the inputs in all spatial directions and increases the number of channels to $256, 512$, and $768$, respectively. The $i$-th downsampling block is composed of a $2$-strided convolution that downsamples the data and increases the number of channels, followed by 4 residual blocks $R_i$,
\begin{equation}\label{eq:resblock}
    R_i(h) = \text{Conv}_3^{256\cdot i}\circ \text{Do}_{0.1}\circ\text{swish}\circ\text{FiLM}(e)\circ\text{GN}\circ\text{Conv}_3^{256\cdot i}\circ\text{swish}\circ\text{GN}(h) + h,
\end{equation}
where $i$ is the downsampling block index, $\text{Do}_{p}$ is dropout with probability $p$, $\text{GN}$ is group normalization, and $\text{FiLM}$ performs feature-wise linear modulation based on a Fourier embedding $e$ of the noise scale $\sigma^t$ with 192 dimensions \citep{perez2017film}. The output of each residual block is denoted $r_{i, j}$, where $i=1, \dots, 3$ is the downsampling block index, and $j=1, \dots, 4$ is the index of the residual block within that resolution level.

\subsubsection{Upsampling stack}

The upsampling stack is composed of 3 upsampling blocks that mirror the downsampling blocks, such that the application of convolutional layers follows the indices $j$ and $i$ in reverse order to the downsampling stack. Each upsampling block applies 4 residual blocks \eqref{eq:resblock} that take as input the sum of the output from the previous block and the output from the symmetrical residual block from the downsampling stack. That is, the output of the $j$-th residual block of the $i$-th upsampling block is defined as
\begin{equation}
    u_{i, j} = R_i(u_{i, j+1} + r_{i, j}), \;\;\; j = 3, \dots, 1.
\end{equation}
The first residual block of the $i$-th upsampling block receives the upsampled output from the $(i+1)$-th level,
\begin{equation}
    u_{i, 4} = R_i(\text{UConv}_3(u_{i+1, 1}) + r_{i, 4}), \;\;\; i=1, 2.
\end{equation}
Here, $\text{UConv}_3$ applies a $3\times 3$ convolution that multiplies the number of channels by $4$, and then reshapes the output to double the spatial degrees of freedom in both spatial directions. The input to the first upsampling residual block is the output from the downsampling stack,
\begin{equation}
    u_{3, 4} = R_i(2r_{3, 4}).
\end{equation}
Finally, the output of the upsampling stack is
\begin{equation}
    u_f = \text{Conv}_3 (\text{UConv}_3(u_{1, 1}) + h_0).
\end{equation}

\subsubsection{Decoding}

The output of the denoiser is finally defined as
\begin{equation}
    D_\theta = \text{Conv}_7\circ \mathcal{I}^{-1} \circ \text{Conv}^6_3 \circ \text{swish} \circ \text{GN} (u_f).
\end{equation}

\subsection{Hyperparameters for model selection}

The hyperparameters used for model selection include the guidance strength $g \in \{-0.1, 0, 0.1, 0.3\}$ used to construct the modified denoiser \eqref{eq:guidance}, the unconditional training probability $p_u \in \{0.05, 0.1, 0.2\}$, the peak learning rate $l_r \in \{5\cdot10^{-5}, 2\cdot10^{-4}\}$, and the batch size $b_s \in \{64, 128, 256\}$. These were optimized in terms of the CRPS of the generative ensembles on the evaluation set.

\section{Diagnostic variables}
\label{sVariables}

The wet-bulb globe temperature (WBGT) is defined in this study following the simplified implementation favored by the Australian Bureau of Meteorology \citep{Willett2012}, since it can be constructed from meteorological variables available in the WUS-D3 dataset. It is defined as
\begin{equation}
    \text{WBGT} = 0.567T+0.393e+3.94 \; ^{\circ}\text{C},
\end{equation}
where $T$ is the air temperature in $^{\circ}$C, and $e$ is the vapor pressure in hPa. The vapor pressure can be readily computed from the specific humidity $q$ and the pressure $p$ as
\begin{equation}\label{eq:vapor_pressure}
    e = \dfrac{pq}{\varepsilon + (1-\varepsilon)q}.
\end{equation}
Here, $\varepsilon\approx 0.622$ is the ratio between the gas constants of dry air and water vapor.

Wildfire risk due to Santa Ana wind events is quantified in terms of the weather component of the Santa Ana wildfire threat index (SAWTI; \cite{Rolinski2016}),
\begin{equation}
    \text{SAWTI} = 0.001w^2(T-T_d),
\end{equation}
where $w$ is the wind speed magnitude at 10 m, and $T_d$ is the dewpoint temperature. The dewpoint temperature is computed from the specific humidity $q$ and the pressure $p$ following Lawrence \citep{Lawrence2005},
\begin{equation}
    T_d = b_1\dfrac{\log({e}/{c_1})}{a_1-\log(e/c_1)}.
\end{equation}
In the previous expression, $a_1=17.625$, $b_1=243.04 ^\circ \text{C}$, and $c_1=610.94$ Pa.

%% file: nc_text/suppl_results_detail.tex
\section{Additional Results}
\label{sAppendixResults}

\subsection{Generative samples}

\begin{figure}[h]
    \centering
    \includegraphics[width=\columnwidth,draft=false]{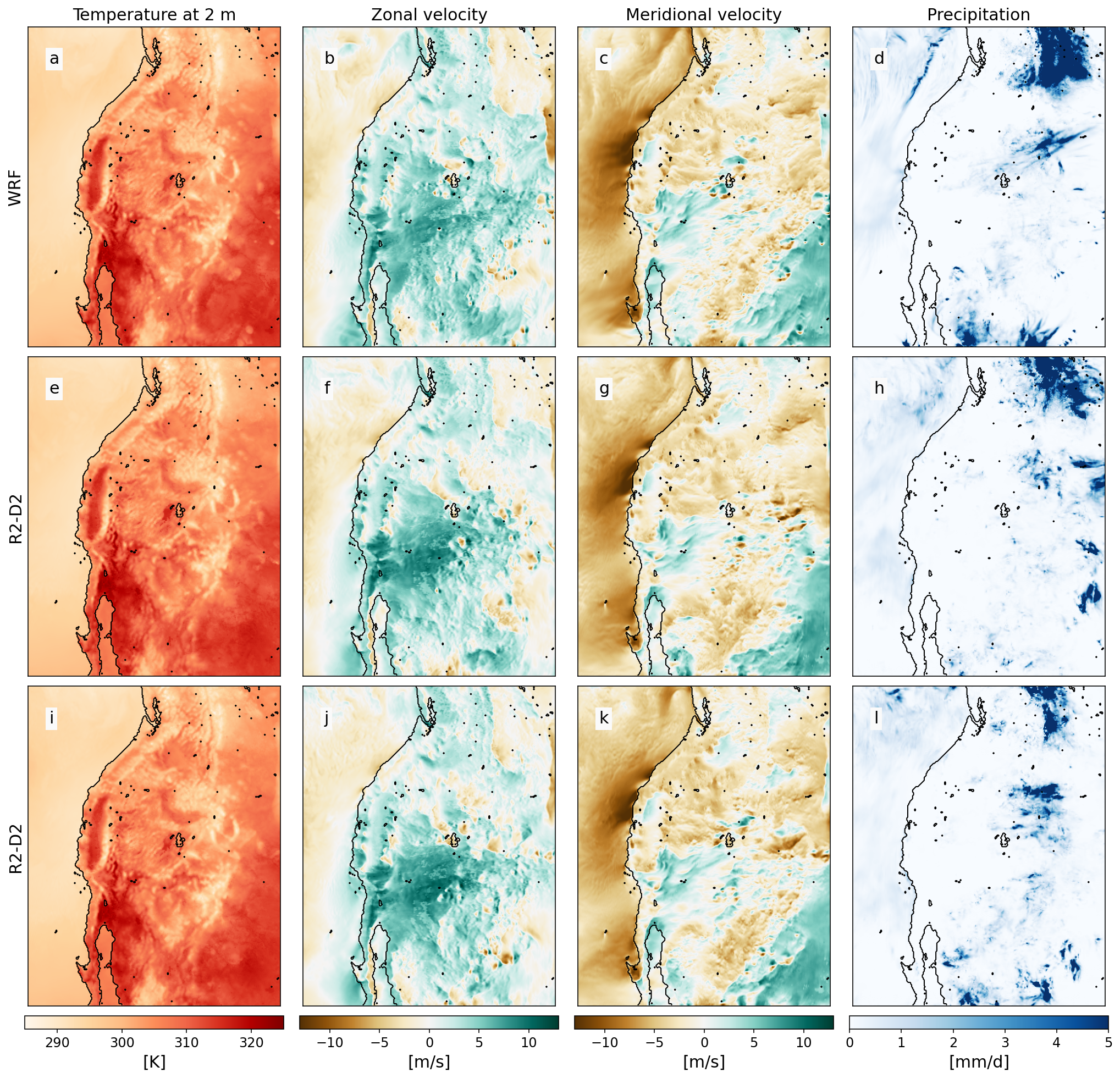}  
    \caption{{Sample downscaled meteorological fields.}
    {Downscaled samples are shown for pure dynamical downscaling with WRF (a-d) and for our dynamical-generative framework (e-l). The date considered is June 30, 2095, 00 UTC, and the forcing ESM is CanESM5. The second and third row show different samples from R2-D2 conditioned on identical inputs.}
    \label{fig:SI_gen_samples}}
\end{figure}

Figure \ref{fig:SI_gen_samples} compares samples generated by R2-D2 with dynamically downscaled snapshots from WRF. The generative samples show similar spatial patterns and effective resolution as the target samples from WRF. In addition, the probabilistic nature of R2-D2 enables it to capture high-resolution variability conditioned on the coarse-scale inputs. The sample variability is most noticeable for precipitation and near-surface winds.

\subsection{Downscaling uncertainty quantification}

Generative downscaling enables capturing the downscaling uncertainty given the large-scale conditioning fields. We can measure how well the generative ensembles capture this uncertainty through the spread-skill ratio \citep{Fortin2014},
\begin{equation}
    r_{ss} = \left[\dfrac{M+1}{M(M-1)}\dfrac{\sum_i^T\sum_m^M(x_{i,m}-\bar{x}_i)^2}{\sum_i^T(\bar{x}_i-o_i)^2}\right]^{1/2},
\end{equation}
where $x_{i,m}$ is generative sample $m$ at time $i$, $o_i$ is the corresponding value downscaled with WRF, $\bar{x}_i$ is the generative ensemble mean, $M$ is the number of ensemble members, and $T$ is the number of dates considered. This ratio, which ideally should be close to unity, summarizes the reliability of the spread of an ensemble prediction system as a measure of downscaling uncertainty.

\begin{figure}[h]
    \centering
    \includegraphics[width=\columnwidth,draft=false]{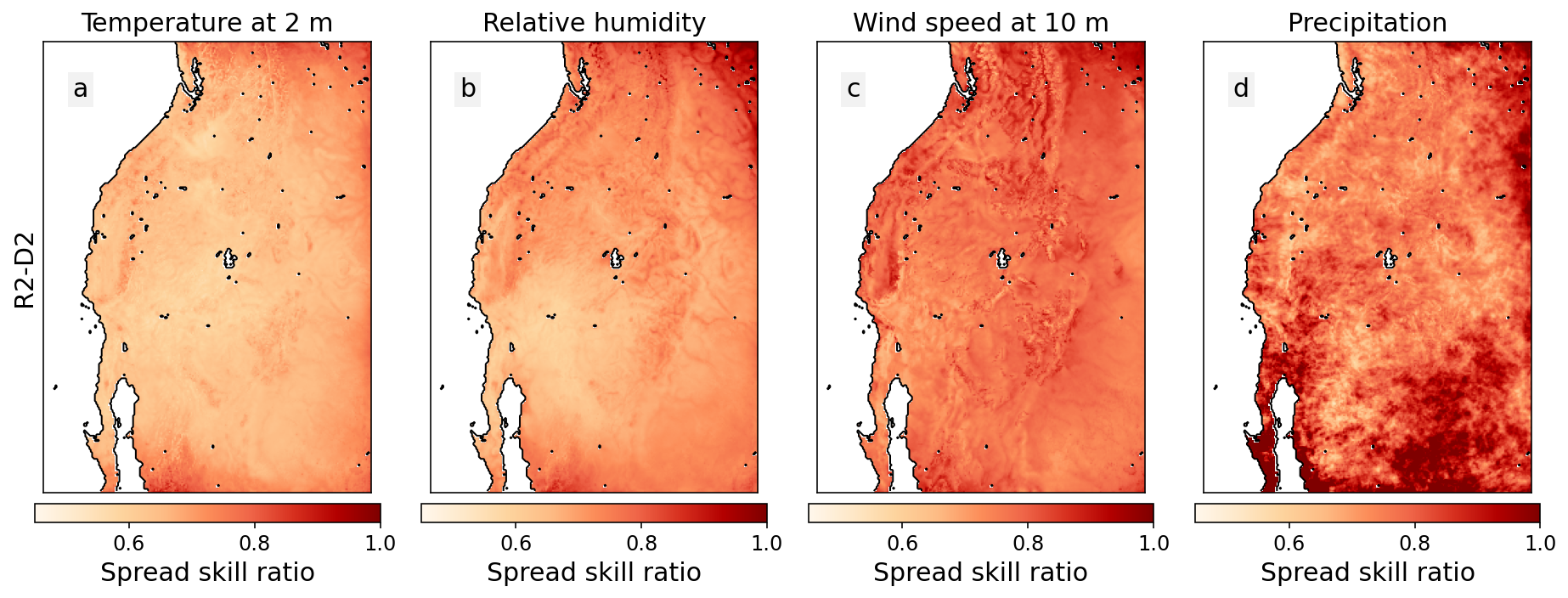}  
    \caption{{Spatial distribution of the spread-skill ratio of the generative ensembles.}
    {Results are computed using 4-hourly data spanning years 2095-2096 of the multi-model SSP3-7.0 climate projection, and using 32-sample ensembles from the generative model R2-D2.}
    \label{fig:SI_UQ}}
\end{figure}

Figure \ref{fig:SI_UQ} explores the reliability of R2-D2 when applied to the full multi-model ensemble. In general, R2-D2 ensembles are slightly underdispersive. Their spread-skill ratio varies by meteorological field, ranging from an average of 0.66 for temperature to 0.80 for precipitation. No robust differences in reliability are observed between modeled and derived variables. As an example, the spread-skill ratio for relative humidity and wind speed are shown Figure \ref{fig:SI_UQ}b,c.

Underdispersion is most noticeable away from the domain boundaries, which couple the coarse and high-resolution trajectories in the data used to train the generative model. We hypothesize that this gradient is due to the type of forcing in the original dataset. More uniform uncertainty estimates may be obtained from generative models trained on dynamically downscaled data nudged toward large-scale conditions from a coarser simulation. This type of forcing is also common in dynamical downscaling frameworks \citep{Rahimi2024}.

\subsection{Downscaling skill}

\subsubsection{Continuous ranked probability score}

\begin{figure}[h]
    \centering
    \includegraphics[width=\columnwidth,draft=false]{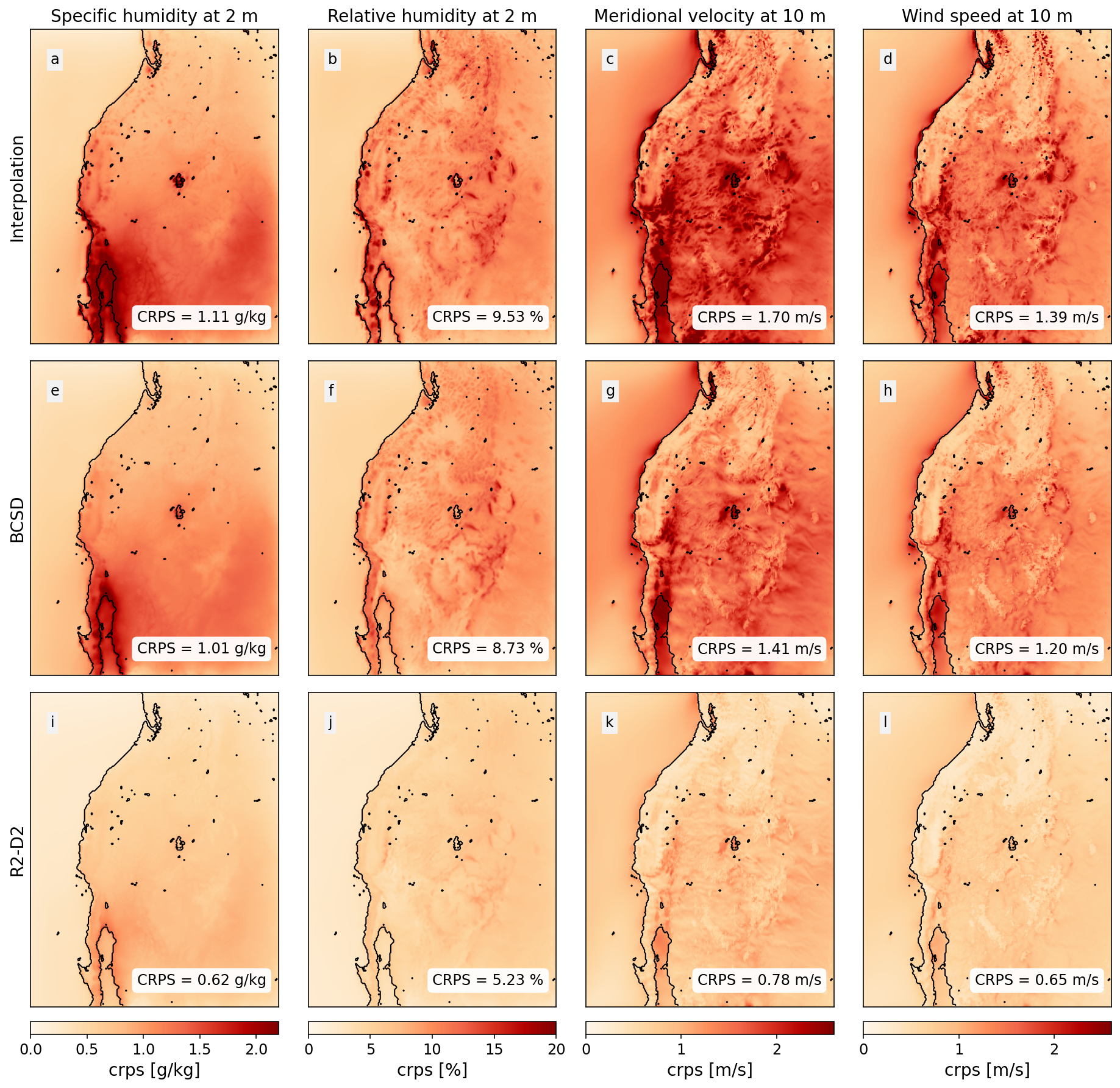}  
    \caption{CRPS of downscaled projections for additional variables.
    {Results are computed using 4-hourly data spanning years 2095-2096 of the multi-model SSP3-7.0 climate projection, and shown for cubic interpolation, BCSD, and the generative model R2-D2. Results for R2-D2 are computed using a 32-member ensemble. The insets show the land-averaged CRPS.}
    \label{fig:SI_add_CRPS}}
\end{figure}

The continuous ranked probability score (CRPS) between an ensemble forecast $x_i$ and an observation $o_i$ is given by
\begin{equation}
    \text{CRPS}(x_i, o_i) = \frac{1}{M} \sum_{m=1}^{M} \vert x_{i,m} - o_i\vert - \frac{1}{2M^2}\sum_{m=1}^{M}\sum_{m'=1}^{M} \vert x_{i,m} - x_{i,m'}\vert.
    \label{eEnsembleCRPSAppendix}
\end{equation}
The results in the main text report the time-averaged CRPS over each spatial location, as well as the spatial average of the time-averaged CRPS over land. The CRPS of additional fields, both modeled and derived, is shown in Figure \ref{fig:SI_add_CRPS}.

\subsubsection{Root mean square error}

\begin{figure}[h]
    \centering
    \includegraphics[width=\columnwidth,draft=false]{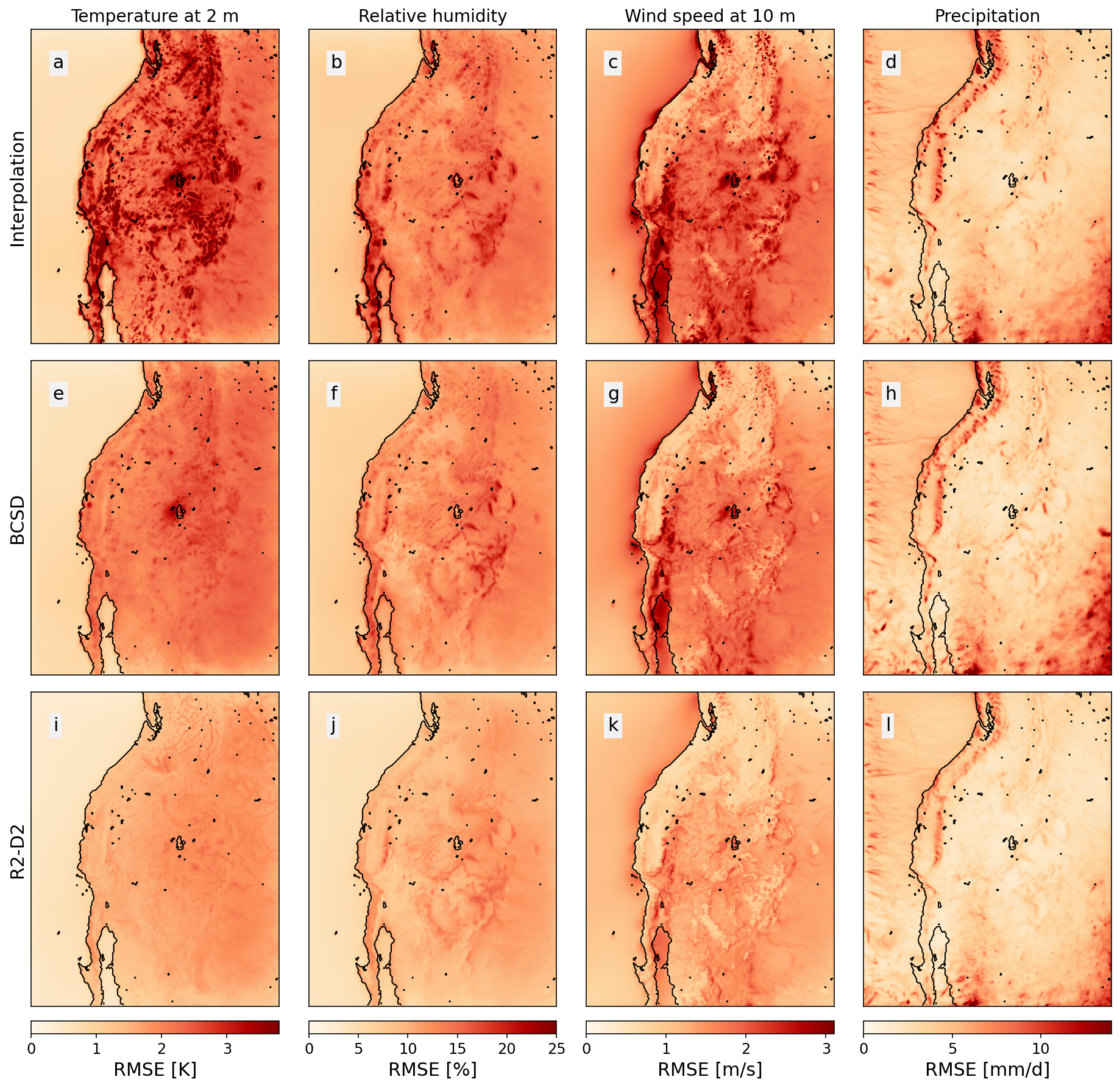}  
    \caption{{Root mean square error of downscaled projections.}
    {Results are computed using 4-hourly data spanning years 2095-2096 of the multi-model SSP3-7.0 climate projection, and shown for cubic interpolation, BCSD, and the generative model R2-D2. Results for R2-D2 are computed using a 32-member ensemble.}
    \label{fig:SI_rmse}}
\end{figure}

\begin{figure}[h]
    \centering
    \includegraphics[width=\columnwidth,draft=false]{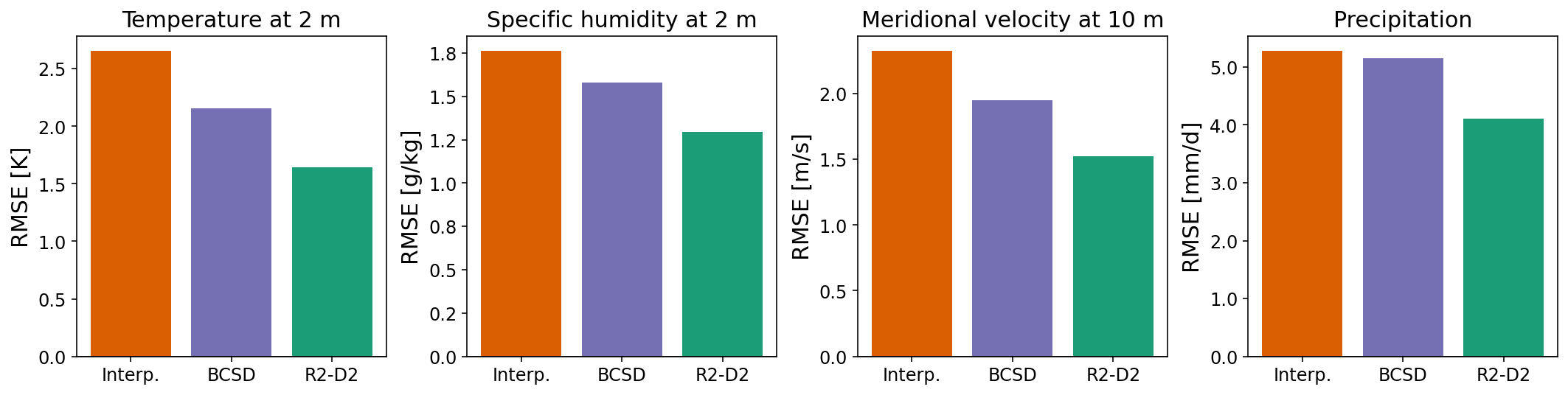}  
    \caption{{Land-averaged downscaling skill measured by RMSE.}
    Results are computed using 4-hourly data spanning years 2095-2096 of the multi-model SSP3-7.0 climate projection, and shown for cubic interpolation, BCSD, and the generative model R2-D2.
    \label{fig:land_avg_RMSE}}
\end{figure}

Figure \ref{fig:SI_rmse} evaluates the downscaling skill for several fields in terms of the root mean square error (RMSE) of the ensemble mean,
\begin{equation}
    \text{RMSE} = \left[\dfrac{1}{T}\sum_i^T{(\bar{x}_i-o_i)^2}\right]^{1/2},
\end{equation}
where $i$ indexes the date of each evaluated snapshot. For the baselines, the downscaled ensemble mean $\bar{x}_i$ simplifies to their deterministic prediction. The mean RMSE over land is also shown in Figure \ref{fig:land_avg_RMSE}.

\subsubsection{Bias}

\begin{figure}[h]
    \centering
    \includegraphics[width=\columnwidth,draft=false]{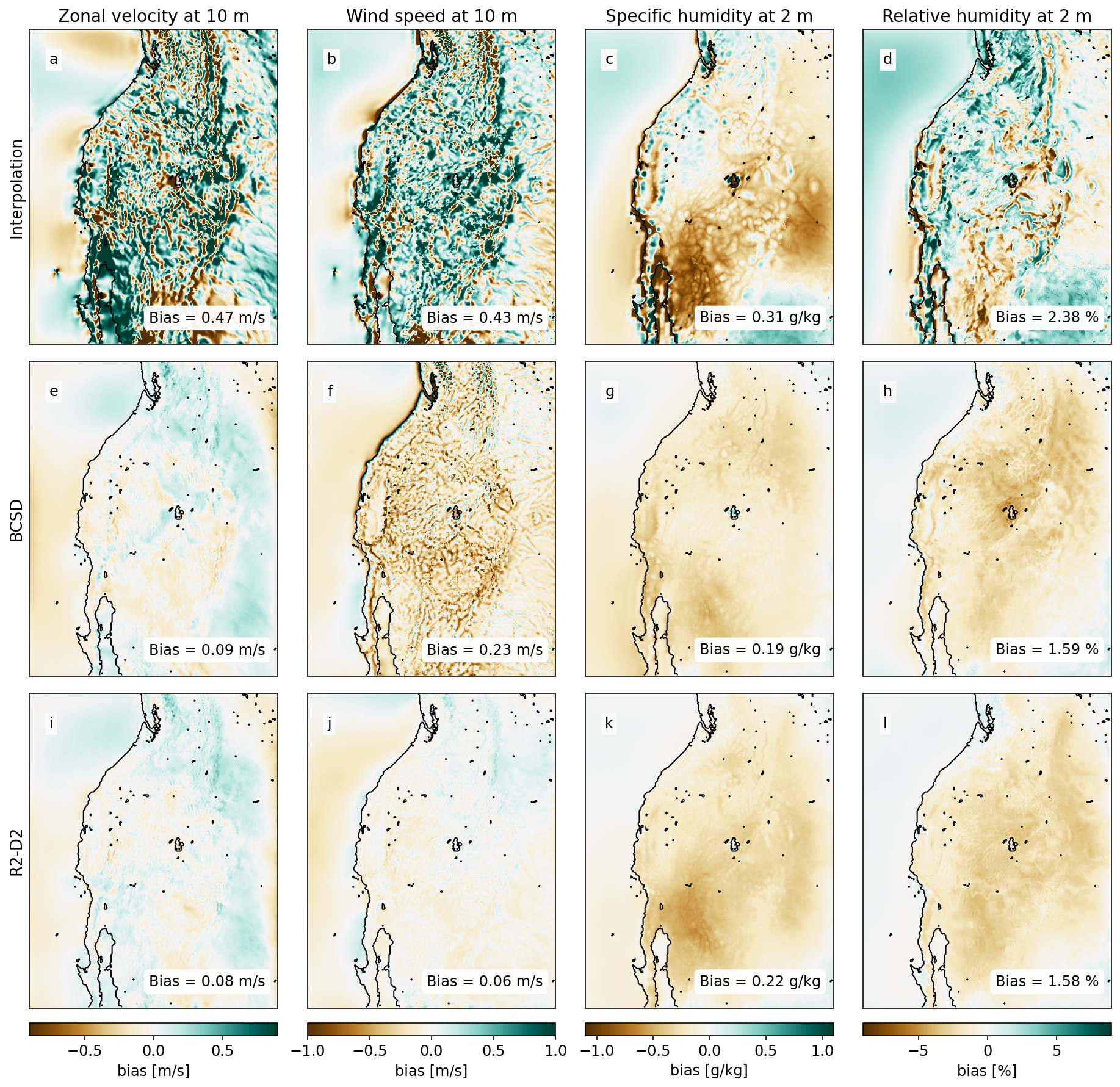}  
    \caption{{Bias of downscaled projections.}
    {Results are computed using 4-hourly data spanning years 2095-2096 of the multi-model SSP3-7.0 climate projection, and shown for cubic interpolation, BCSD, and the generative model R2-D2. Results for R2-D2 are computed using a 32-member ensemble. The insets show the land-averaged mean absolute bias.}
    \label{fig:SI_bias}}
\end{figure}

The bias of a set of ensemble forecasts $x=\{x_i\}$ with respect to a set of observations $o=\{o_i\}$ covering the same period is given by
\begin{equation}
    \text{Bias}(x,o) = \frac{1}{T}\sum_{i=1}^{T}\left[\frac{1}{M} \sum_{m=1}^{M} x_{i,m} - o_{i}\right].
    \label{eq:BiasAppendix}
\end{equation}
The results in the main text report the bias over each spatial location, computed over years 2095 and 2096 and for all 8 forcing ESMs. The bias of additional fields is shown in Figure \ref{fig:SI_bias}. The generative model yields samples with a similar reduction in bias in directly modeled and derived variables. Examples of the latter are the relative humidity and the wind speed magnitude. Univariate quantile-based methods such as BCSD tend to yield biased samples for derived variables, due to their inability to capture multivariate correlations. This is most noticeable for the near-surface wind speed, shown in Figure \ref{fig:SI_bias}f.

\subsubsection{ESM learning sensitivity}

\begin{figure}[h]
    \centering
    \includegraphics[width=\columnwidth,draft=false]{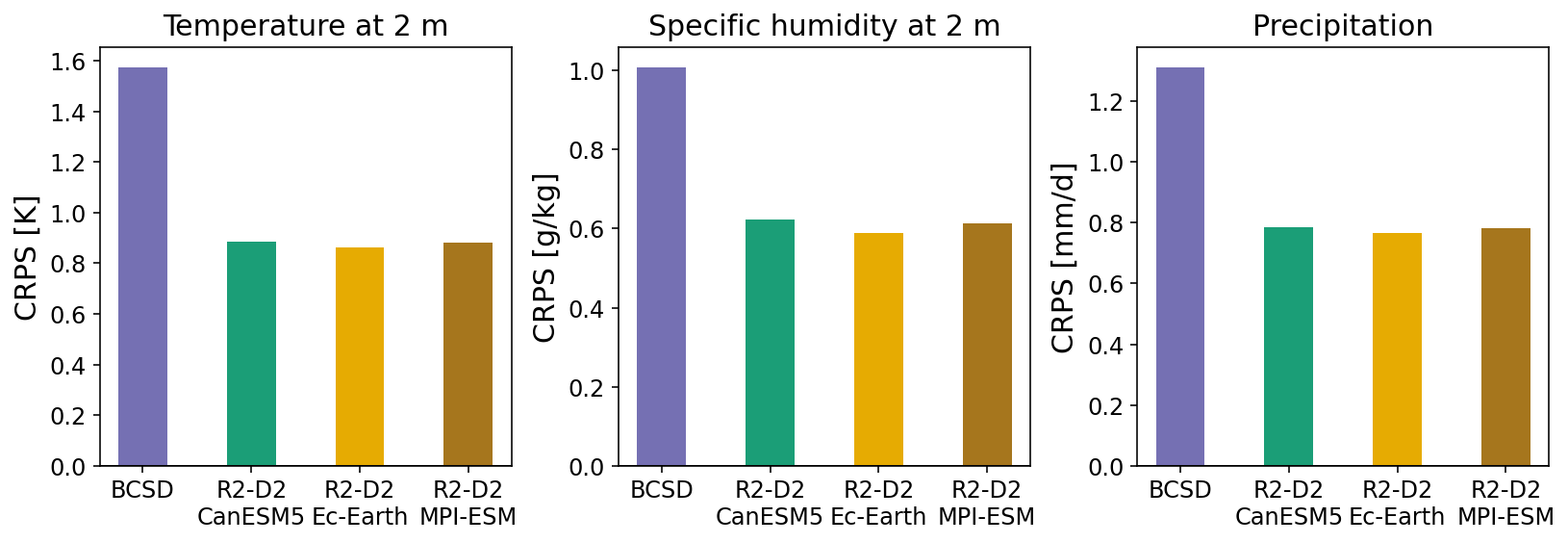}  
    \caption{{Land-averaged downscaling skill of generative models trained on different ESMs as measured by CRPS.}
    Results are computed using 4-hourly data from 2095 and 2096 of the multi-model SSP3-7.0 climate projection, and shown for BCSD, and the generative model R2-D2 trained using data from three different forcing ESMs: CanESM5, EC-Earth3-Veg, and MPI-ESM1-2-HR.
    \label{fig:model_sens_crps}}
\end{figure}

The results shown in the main manuscript evaluate the skill of an R2-D2 model trained on dynamically downscaled data from CanESM5. To demonstrate that the conclusions drawn are not sensitive to the ESM used to train R2-D2, the CRPS of the downscaled multi-model projections are recomputed in Figure \ref{fig:model_sens_crps} for models trained on data from EC-Earth3-Veg and MPI-ESM1-2-HR. The observed differences are much smaller than the differences with respect to the BCSD baseline, and similar to those that can be expected from training different generative models on the same data, due to the stochastic nature of the learning process. This suggests that the generative stage of the dynamical-generative framework can be trained on any model for which downscaled results are available, without performance degradation.

\subsection{Multi-model projection analysis}

\subsubsection{Metrics}

The reliability of the downscaled multi-model projections is assessed in terms of the ability to match the value of the quantiles of the target distribution, which in this case is the dynamically downscaled multi-model distribution.

Given a time period $\tau$, a sampling frequency $f$, a number of forcing ESMs $N=8$, and a number of samples per input $M$, we compute the quantiles associated with the distribution of the $MN\tau f$ samples for each location. For the deterministic baselines and dynamical downscaling, $M=1$. We leverage the probabilistic nature of R2-D2 to draw $M=${32} samples per input. We aggregate 3-month periods with daily sampling frequency, such that the number of samples used for the deterministic systems is $720$, and $2.3\cdot10^4$ for the dynamical-generative framework. In all cases, we evaluate the sampling uncertainty by drawing $100$ bootstrapped estimates of the quantiles and computing their standard deviation.

Let $x_q$ be the estimated value of the quantile $q$ such that $\text{Pr}[X \leq x_q] \geq q$. The mean absolute error (MAE) of quantile $q$ is defined as
\begin{equation}
    \text{MAE}_q = \vert x_q - o_q\vert,
\end{equation}
where $x_q$ is the quantile estimate obtained from a downscaling framework, and $o_q$ is the target quantile value, which in this case is taken to be the quantile from the full dynamically downscaled ensemble. The land-averaged quantile MAE shown in Figure \ref{fig:quantiles} is simply the spatial average of this error over land. In addition, the reliability of the downscaled projections is evaluated locally by plotting the projected quantiles as a function of the target quantiles. A perfectly reliable system would follow the $y=x$ diagonal.

\subsubsection{Additional variables and locations}

\begin{figure}[h]
    \centering
    \includegraphics[width=0.9\columnwidth,draft=false]{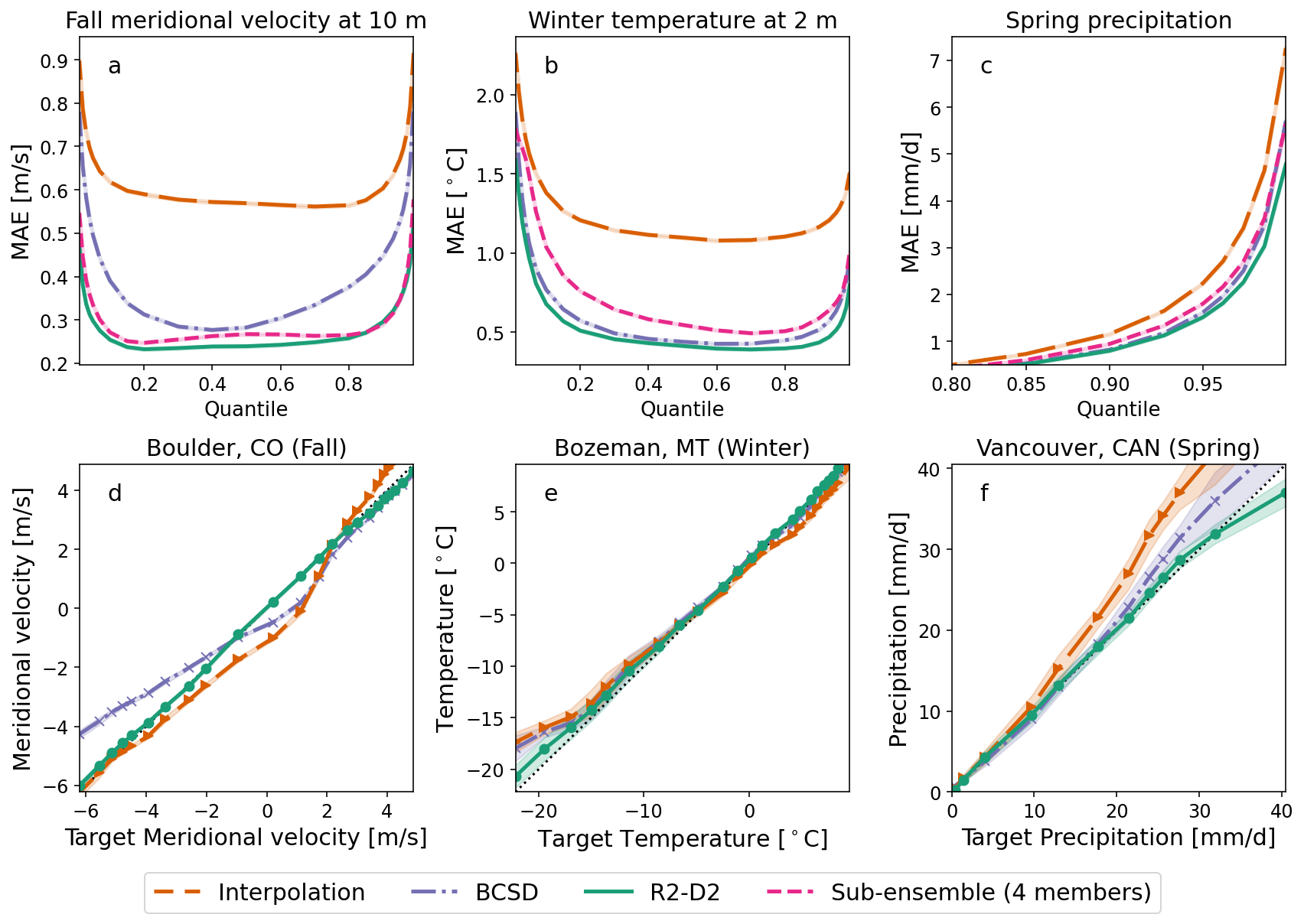}  
    \caption{As in Figure \ref{fig:quantiles} but for fall meridional wind velocity at 10 m, winter temperature at 2 m, and spring precipitation. Local quantile-quantile plots (d-f) are shown for the cities of Boulder (Colorado), Bozeman (Montana), and Vancouver (Canada).}
    \label{fig:SI_quantiles}
\end{figure}

Figure \ref{fig:SI_quantiles} evaluates the multi-model distributions projected by different downscaling methods for additional variables and locations. Local results are shown for the high-elevation cities of Boulder (Colorado) and Bozeman (Montana), as well as coastal Vancouver (Canada). R2-D2 is able to predict the probability of winter cold spells better than the baselines near Bozeman, which sits in the Gallatin Valley surrounded by the Rocky Mountains. Both interpolation and BCSD overestimate the $1\%$-quantile temperature by more than $3^\circ$C.  Dynamical-generative downscaling is also superior in the prediction of equatorward winds in Boulder, Colorado. During the spring, its skill is comparable to BCSD in the prediction of precipitation over coastal Vancouver.

\subsubsection{Generalization gap across forcing ESMs}

\begin{figure}[h]
    \centering
    \includegraphics[width=0.9\columnwidth,draft=false]{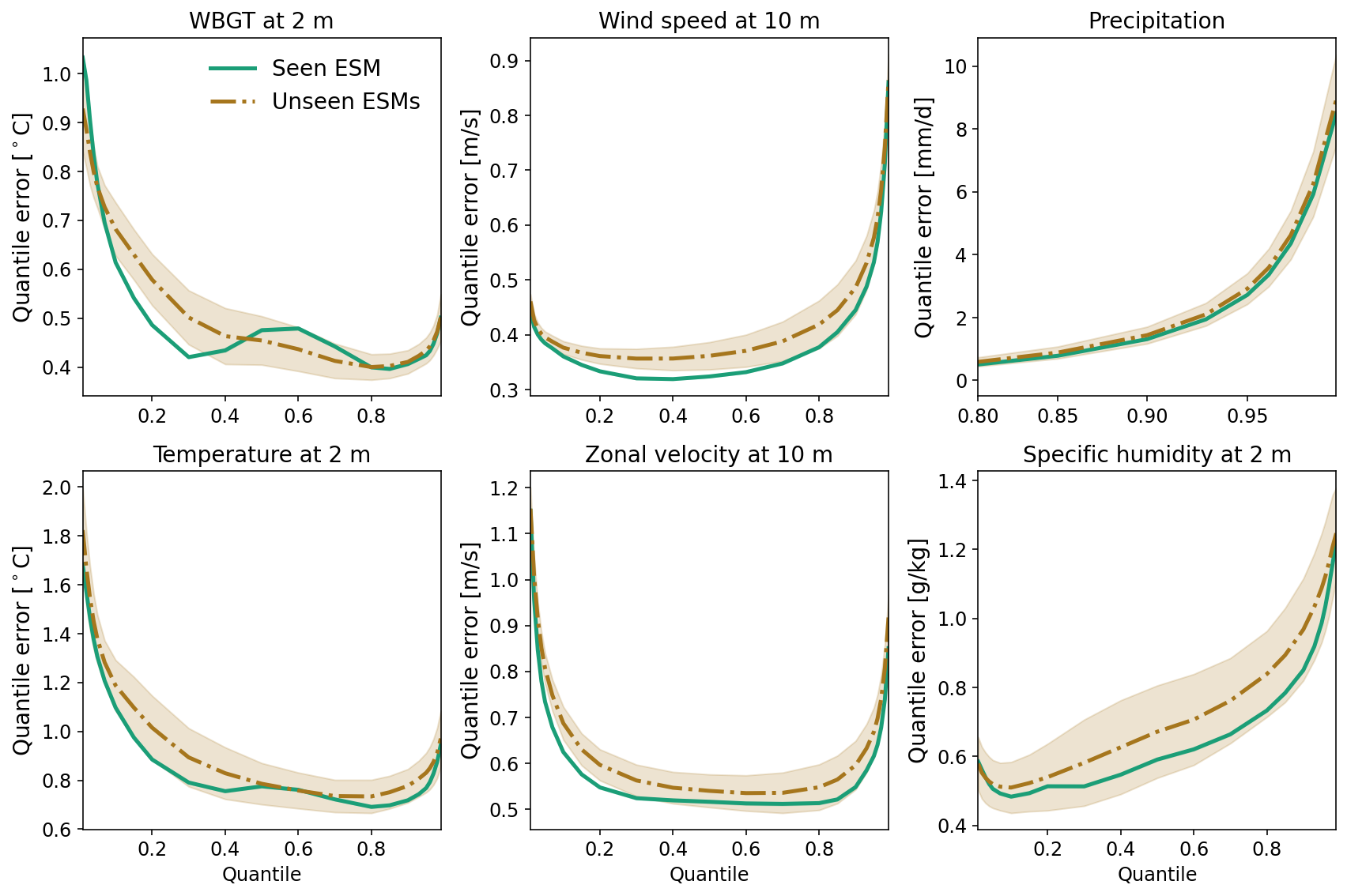}  
    \caption{Mean absolute error of downscaled quantiles over land, with respect to the corresponding dynamically downscaled quantiles. Results are shown for inputs from the ESM used to train the generative model, CanESM5 (Seen ESM), and for the other 7 ESMs not seen during training (Unseen ESMs).
    Quantiles are computed using daily snapshots at 00 UTC for the June-August period of 2095. Uncertainty estimates represent the bootstrapped standard deviation.}
    \label{fig:SI_gengap_quantiles}
\end{figure}

Figure \ref{fig:SI_gengap_quantiles} explores the generalization gap when dynamical-generative downscaling is applied to ESMs unseen during training, in terms of the ability to capture the quantiles of their target downscaled distributions. We compute the MAE of downscaled quantiles for unseen ESM projections, and compare them against the MAE of downscaled quantiles for projections performed with the ESM seen during training (CanESM5). The quantile errors for CanESM5 projections are lower than average, but the spread across ESMs is comparable to this generalization gap.

\subsection{Additional samples for Santa Ana wind event}

\begin{figure}[h]
    \centering
    \includegraphics[width=0.9\columnwidth,draft=false]{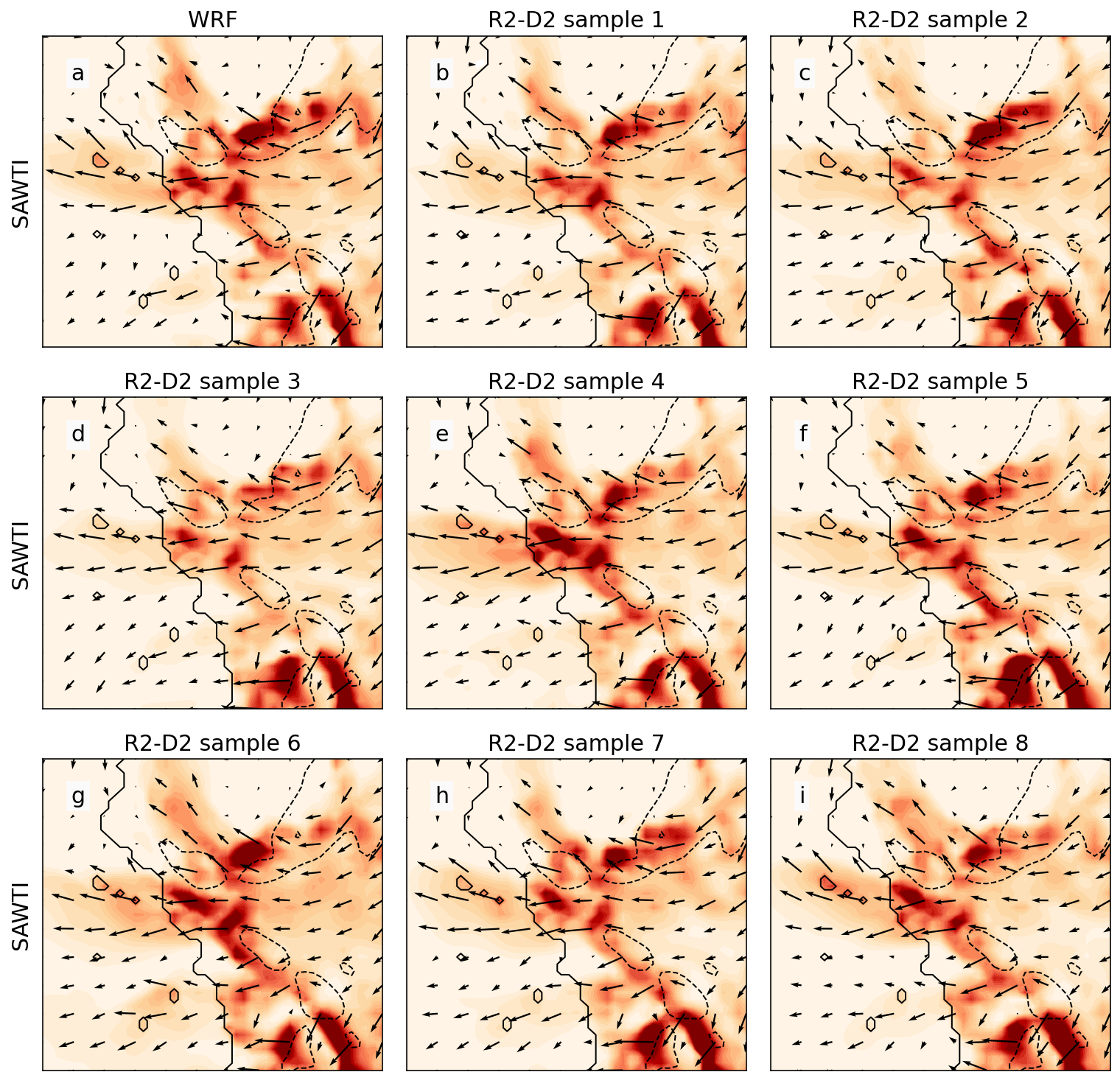}  
    \caption{Downscaled projections of the Santa Ana wind event of November 13, 2095, generated with WRF (a), and R2-D2 (b-i). The color represents the magnitude of the Santa Ana wildfire threat index, and the arrows represent the magnitude and direction of near-surface winds. The conditioning inputs are constant and taken from the EC-Earth3-Veg SSP3-7.0 projection, at 00 UTC.}
    \label{fig:santa_ana_samples}
\end{figure}

Figure \ref{fig:santa_ana_samples} depicts additional generative downscaled samples from R2-D2 conditioned on the Santa Ana wind event conditions described in the main text. Some variability is observed in the magnitude of SAWTI, but overall the wildfire risk conditions are similar in all generated samples.

